\documentclass[aps,floats,twocolumn,epsf,prb,showpacs]{revtex4-1}
\usepackage{graphicx}
\usepackage{epstopdf}
\usepackage{amsmath}
\usepackage{amssymb}
\usepackage{booktabs}
\usepackage[colorlinks=true,urlcolor=blue,citecolor=blue,linkcolor=blue,breaklinks=true]{hyperref}
\usepackage{color}
\usepackage[dvipsnames]{xcolor}

\newcommand{\orcidlink}[1]{}

\begin{document}

\title{Pb$_{10-x}$Cu$_x$(PO$_4$)$_6$O: a Mott or charge transfer insulator \\in need of further doping for (super)conductivity}

\author{Liang Si$^{a,b}$\orcidlink{0000-0003-4709-6882}, Markus Wallerberger$^b$,  Andriy  Smolyanyuk$^b$,  Simone di Cataldo$^b$, Jan M. Tomczak$^{b,c}$, and Karsten Held$^b$\orcidlink{0000-0001-5984-8549}}
\affiliation{$^a$ School of Physics, Northwest University, Xi'an 710127, China\\$^b$ Institute of Solid State Physics, TU Wien, 1040 Vienna, Austria\\
$^c$ Department of Physics, King’s College London, Strand, London WC2R 2LS, United Kingdom }
  

\date{\today}

\begin{abstract}
  We briefly review the status quo of research on the putative superconductor Pb$_9$Cu(PO$_4$)$_6$O also known as LK-99.  Further, we provide {\em ab initio} derived tight-binding parameters for a two- and five-band model, and solve these in dynamical-mean-field theory. 
  The ratio interaction-to-bandwidth makes  LK-99 a Mott or charge transfer insulator. Electron or hole doping (which is different from substituting Pb by Cu and thus differs from LK-99) is required to make it metallic and potentially superconducting.
  \end{abstract}

\maketitle

\section{ Introduction}
In recent preprints~\cite{Lee2023_2,Lee2023_3}, Lee, Kim, {\em et al.} reported the discovery of a {room-temperature superconductor
at  ambient-pressure:} Pb$_{10-x}$Cu$_x$(PO$_4$)$_6$O with $0.9<x<1.1$. They {had} previously named this material LK-99
after their initials and the year of the first synthesis.
Their more recent samples show somewhat stronger signatures of superconductivity~\cite{Lee2023_1,Lee2023_2,Lee2023_3}: (i) a sharp drop in the resistivity~\cite{Lee2023_1,Lee2023_3}, according to Ref.~\onlinecite{Lee2023_2} to the order of $10^{-10} - 10^{-11}\,\Omega$cm though in Refs.~\onlinecite{Lee2023_1,Lee2023_3} a higher noise level is visible, (ii) a negative (diamagnetic) spin susceptibility and levitation on a magnet~\cite{Lee2023_3}, and (iii) sharp voltage jumps at critical currents, with the critical  currents vanishing in approximately a quarter-circle as a function of  temperature and magnetic field \cite{Lee2023_1,Lee2023_2,Lee2023_2}.

If LK-99 is truly a superconductor at {ambient} temperature and pressure, it is arguably one of the most significant physics discoveries of recent history. However, experimental confirmation is urgently needed{: The} above experiments, while indicative of superconductivity, do not unambiguously prove it. (i) The noise level of the resistivity appears too large for concluding that LK-99 has zero resistance.  (ii) The negative susceptibility and levitation can be caused by a simple diamagnet. (iii) The voltage jumps might also be caused by contact issues. One has to admit however that taken everything together,  the overall picture provides quite some indication for superconductivity. If the critical temperature was  1~K [and if taking into account the first confirmations of (i) and (ii), see below],  the scientific community would now most likely be quite positive that at least parts of the  LK-99 sample are superconducting. But room temperature superconductivity is an extraordinary claim, and extraordinary claims rightfully require an extraordinarily solid proof. Such waterproof evidence has not been given  as of the time of submitting this article, neither has solid evidence against.

Naturally, the results by Lee, Kim {\em et al.} led to huge experimental and theoretical efforts. Let us briefly review the status quo of these subsequent works as of the day of submission (Aug. {8}th, 2023):
The levitation (i) has been reproduced by Wu {\em et al.} \cite{Wu2023} and further groups on social media. A sharp drop in resistivity has been confirmed by Hou {\em et al.}~\cite{Hou2023}, albeit at 100\,K instead of 100$^\circ$ Celsius.  Hou {\em et al.} also report two strange resistivity jumps above 250\,K
(which the authors suggest might  be caused by issues with the electrode contacts) as well as an abnormal field dependence.

In contrast, other experimental groups report an opposite behavior. 
Liu~{\em et al.}~\cite{Liu2023} find an increase of the resistivity with decreasing temperature indicating that LK-99 is a semiconductor or insulator. They also observe a paramagnetic response instead of a diamagnetic one, and the magnetic susceptibility {as well as} the resistivity increases with decreasing temperature.
Kumar~{\em et al.}~\cite{Kumar2023,Kumar2023_2} also successfully synthesized Pb$_{10-x}$Cu$_x$(PO$_4$)$_6$O in a modified lead apatite structure, {and report a diamagnetic insulator.
Abramiam {\em et al.}~\cite{Abramian2023} conjecture that the samples by Lee {\em et al.} are not pure LK-99, but a coexistence of superconducting and non-superconducting regions; superconductivity might emerge from  another material.
Guo {\em et al.} \cite{Guo2023} report a ferromagnetic hysteresis and half levitation.}

As for theory, density functional theory (DFT)\cite{PhysRev.136.B864} is state-of-the-art for calculating  crystal structures and for getting, at the bare minimum, a first idea of the electronic structure. 
Five groups~\cite{Lai2023,Griffin2023,Si2023,Kurleto2023,CabezasEscares2023} independently performed such DFT calculations, appearing on arXiv within days, and showing similar results (cf.~\cite{Kumar2023_2,Tao2023,Sun2023}): For the lead apatite crystal structure with one Pb atom replaced by Cu, two very flat bands cross the Fermi energy. Below these are still flat, but slightly more dispersive O bands, and another Cu band.
Some of the DFT calculations also analyze possible alternative Cu and O positions~\cite{Si2023,Griffin2023,CabezasEscares2023,Sun2023} so far to a very limited extent; Cabezas-Escares\cite{CabezasEscares2023}  find an instability with a simplified frozen phonon calculation.
Lai~{\em et al.}~\cite{Lai2023} suggest that gold-doped lead apatite may have stronger effects than Cu.
Griffin~\cite{Griffin2023}, Si and Held~\cite{Si2023}, and Kurleta~{\em et al.}~\cite{Kurleto2023}  argue that the flat bands might boost electron-phonon mediated superconductivity;  Si and Held~\cite{Si2023}  also suggest purely electronic flat-band superconductivity~\cite{Kuroki2005,Iglovikov2014,Aoki2020} as a possible alternative. 

\begin{figure}[tb]
\includegraphics[trim = 22cm 0cm 4cm 0cm, width=.6\linewidth,angle=0,clip]{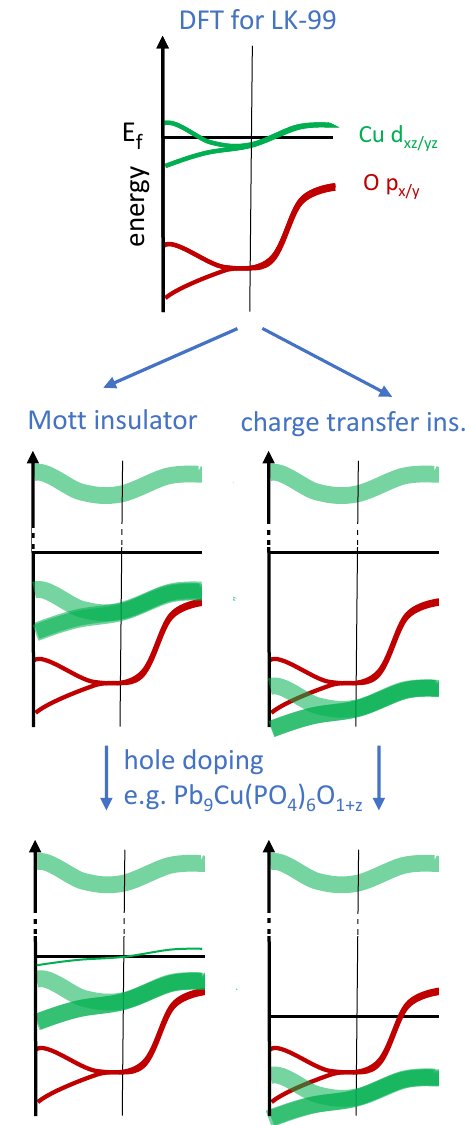}
\caption{Top: Schematics of DFT bandstructure for  Pb$_{9}$Cu(PO$_4$)$_6$O.  Middle: Mott-Hubbard splitting of the Cu $d_{xz,yz}$ orbitals leading to a Mott or charge transfer insulator. Note here we visualize the case with an orbital symmetry breaking (ordering) indicating by having different orbitals in the lower and upper Hubbard band. Interrupted lines indicate a larger energy separation. Bottom: if doped the  Mott or charge transfer insulator becomes metallic. here hole doping is visualized. For electron doping the quasiparticle band (and Fermi energy) would be closer to the upper Hubbard band. A charge transfer insulator {for electron doping is not plausible,} because the next unoccupied orbitals (Pb $p$) are too high in energy~\cite{Si2023}.
        \label{Fig:sketch}}
\end{figure}

Some groups~\cite{Lai2023,Griffin2023,Kurleto2023,CabezasEscares2023} consider the flat DFT bands crossing the Fermi energy as evidence that Cu doping $x\approx1$ makes {insulating} lead apatite metallic{, thus} explaining the conducting and prospectively superconducting state of LK-99.
While this is suggestive from the  DFT results, Si and Held~\cite{Si2023} 
estimate the interaction-to-bandwidth $U/W$ to be of ${\cal O}(10)$ and thus conclude that LK-99 must be a Mott (or charge transfer) insulator,
see Fig.~\ref{Fig:sketch} for an illustration.
They further conjecture that the accompanying spin-1/2 should show a strong paramagnetic response so that a diamagnet without  superconductivity  is difficult to imagine.

Indeed such a {Mott\cite{Gebhard1997} or charge transfer~\cite{Zaanen1985}} insulator {might} explain the simultaneous experimental findings of a paramagnetic insulator. At the same time, the  metallic (and prospectively superconducting) behavior found in the other experiments~\cite{Lee2023_1,Lee2023_2,Lee2023_3,Wu2023,Hou2023} is possible if {(part of)} the sample is doped, see Fig.~\ref{Fig:sketch} (bottom).   Indeed high-temperature cuprate superconductors~\cite{Bednorz1986} are in the same class of a charge transfer (or Mott) insulator~\cite{Scalapino2012}, and also have Cu and O orbitals as the relevant ones. However, this is how far the similarity goes. {Superconducting cuprates have $U/W$  only of ${\cal O}(1)$,} and the Cu atoms form a square lattice in the CuO$_2$ planes, while the lead apatite crystal structure is hexagonal.
To  make such a Mott or charge transfer insulator metallic, one needs electron or hole doping  which is not
possible by changing $x$, i.e., the ratio of Cu:Pb. 
The synthesis procedure~\cite{Lee2023_3} and lead apatite crystal structure suggests that  electron doping was possible for $0 < y \ll 1$ and $z<0$ and hole doping for $z>0$ in
Pb$_{10-x}$Cu$_x$(P$_{1-y}$S$_y$O$_4$)$_6$O$_{1+z}$.  Note,  the nominal oxidation states {are:} Pb$^{2+}$,  Cu$^{2+}$,  P$^{5+}$, S$^{6+}$ and O$^{2-}$. Anyhow, this is merely an educated guess.

Also based on the picture of a doped Mott insulator, Baskaran~\cite{Baskaran2023} speculates that  Cu atoms cluster in chains or 2D
patches with a Cu$^0$ electronic configuration instead of Cu$^{2+}$ \cite{Lai2023,Griffin2023,Si2023,Kurleto2023,CabezasEscares2023}, and thus realize his theory
of a broad band Mott localization.
First groups also started doing calculations for two-orbital models on a triangular lattice, using a  Bardeen, Cooper and Schrieffer(BCS)~\cite{BCS1957} type of coupling~\cite{Tavakol2023} and slave-boson mean field theory. The authors find  $f$-wave and $s$-wave superconductivity (though at too low critical temperatures)~\cite{Oh2023}, respectively.

\begin{figure*}[t]
\centering
\includegraphics[width=0.95\textwidth]{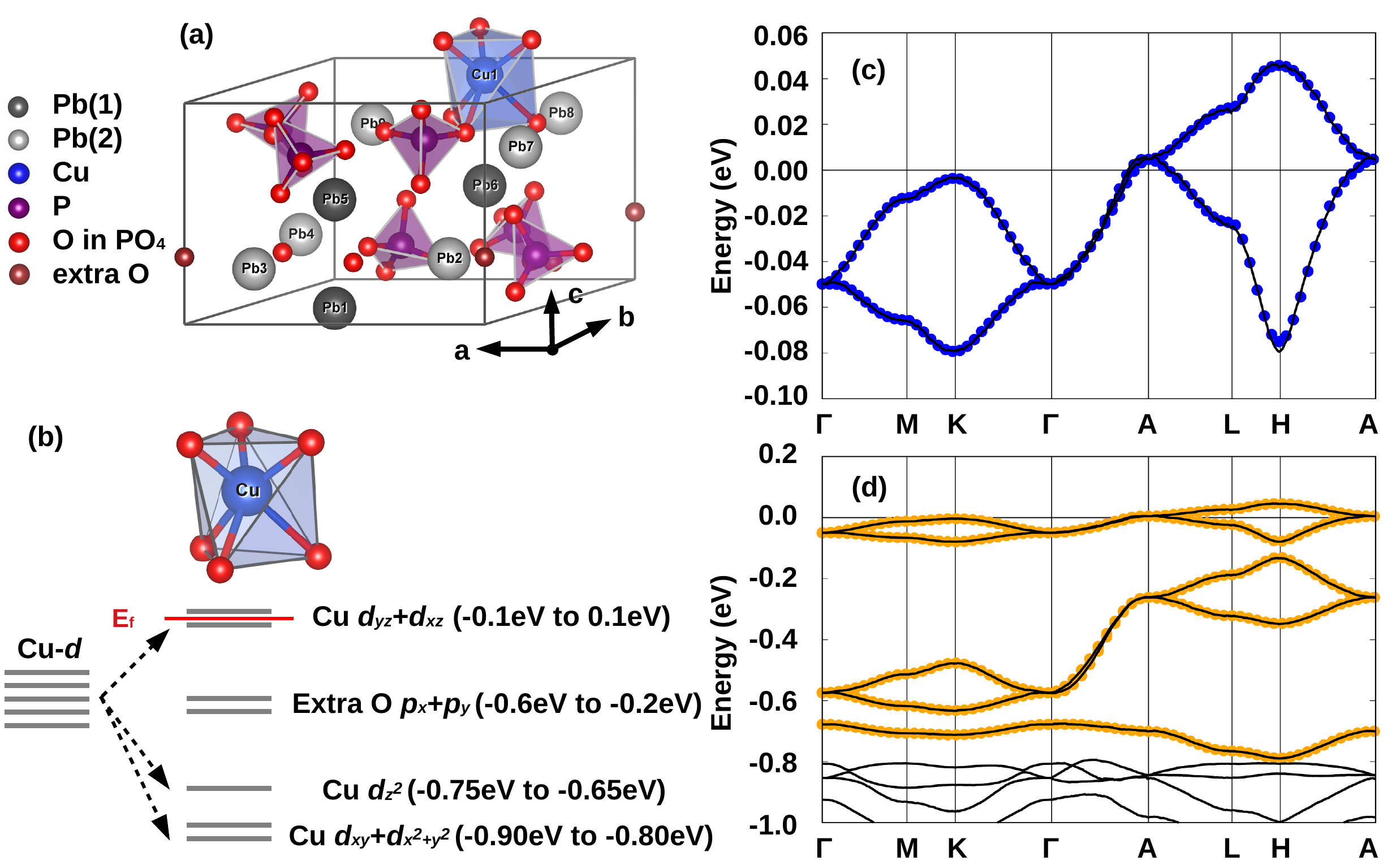}
\caption{(a) DFT-relaxed structure of Pb$_9$Cu(PO$_4$)$_6$O; (b) schematic figure of energy band splitting 
from the octahedral Cu-coordination of the CuO$_6$ motif;
Wannier projections  for the two bands model (c) and five bands model (d), respectively. The Wannier bands (dots) are virtually identical to the DFT bands (lines).}
\label{Fig1_structure}
\end{figure*}

In this paper, we would like to put such tight binding parameters on a more solid basis. To this end, we {do a Wannier function projection} and calculate the tight binding parameters for: {(i)} a two-band low energy model made up of the  Cu $d_{xz/yz}$ orbitals {and (ii)}
a five-band model also involving the O $p_{x/y}$ states just below these and{, additionally,} the next Cu {($d_{z^2}$)} orbital, see   Fig.~\ref{Fig:sketch} (top) {[this Fig.~does not include the 3rd Cu $d$ orbital below the O $p_{x/y}$ bands, {cf.~Fig.~\ref{Fig1_structure} below}]}.
{These tight-binding models can be used for subsequent many-body calculations and are listed in \ref{Tab1_2bands} and \ref{Tab2_5bands}, respectively.}
We motivate the strength of the  Kanamori interaction parameters on the Cu sites. Finally we solve these models in DMFT and find a Mott or charge transfer insulator. A similar insulator is also obtained in DFT+$U$, but only if the crystal symmetry lifts the degeneracy of the  Cu $d_{xz/yz}$ orbitals.

Let us put some caveats here  regarding the low-energy model. It assumes the periodic continuation of a unit cell with a single formula unit ($x=1$, $y=0$, $z=0$) {and} optimized O and Cu positions~\cite{Si2023}. This yields among others, a regular triangular lattice of the Cu sites.
Other O and Cu positions are{, however,} so close in energy~\cite{Si2023} that we must expect a disordered arrangement of these at room temperature {--- unless there is a} crystal distortion stabilizing some arrangement. The x-ray diffraction (XRD) {patterns}~\cite{Lee2023_3,Liu2023,Hou2023} clearly indicate an undistorted lead apatite structure without periodic arrangement of the Cu atoms. For the matter of Mott insulator or not, this is not relevant, but for (super)conductivity the possible long-range ordering of the Cu and O atoms or vice versa a disordered arrangement of these is very relevant.

\begin{table*}[tbh]
\begin{tabular}{cc|rrrrrrrrrrrrrr}
\toprule
Orbital-1 & Orbital-2& $t_{000}$ & $t_{1}$ & $t_{1}$ & $t_{1}$ & $t_{1}$ & $t_{1}$ & $t_{1}$ &$t_{2}$ & $t_{2}$ & $t_{2}$ & $t_{2}$ & $t_{2}$ & $t_{2}$ & $t_z$ \\
 &  & (000) & (100) & (010) & (110) & (-1-10) & (-100) & (0-10) & (120) & (210) & (1-10) & (-1-20) & (-2-10) & (-110) & (001) \\
\midrule
$e_g$(2)  & $e_g$(2) &  $-24$ & $  6$ & $ -5$ & $ 0$ & $ 0$ & $  6$ & $ -5$ & 0 & 0 & 0 & 0 & 0 & 0 & $-10$ \\
$e_g$(1)  & $e_g$(1) &  $-24$ & $ -6$ & $  5$ & $ 0$ & $ 0$ & $ -6$ & $  5$ & 0 & 0 & 0 & 0 & 0 & 0 & $-10$ \\
$e_g$(2)  & $e_g$(1) &  $  0$ & $  6$ & $  4$ & $-2$ & $15$ & $-11$ & $-13$ & 0 & 0 & 0 & 0 & 0 & 0 & $  0$ \\
$e_g$(1)  & $e_g$(2) &  $  0$ & $-11$ & $-13$ & $15$ & $-2$ & $  6$ & $  4$ & 0 & 0 & 0 & 0 & 0 & 0 & $  0$ \\
\bottomrule
\end{tabular}
\caption{Two-band model, comprising two Cu $e_g$(1) and $e_g$(2) orbitals (corresponding to $d_{yz}$, $d_{yz}$). Hopping parameters up to second nearest from Cu to Cu sites and Orbital-1 to Orbital-2. Here, $t_1$ and $t_2$ denote 1st and 2nd nearest hoppings. The numbers in brackets indicate the hopping vector in real space,
see Fig.~\ref{Fig3_orbchar}(f).
Hopping terms that are close to or smaller than 1\,meV are put as 0\,meV. All hoppings are in units of meV; $t_{000}$ is the on-site energy.
\label{Tab1_2bands}}
\end{table*}

\begin{table}[tbh]
\begin{tabular}{cc|rrrrrrrr}
\toprule
Orb-1   & Orb-2 & $t_{000}$ & $t_{1}$ & $t_{1}$ & $t_{1}$ & $t_{1}$ & $t_{1}$ & $t_{1}$ & $t_z$ \\
 & & {\scriptsize(000)} & {\scriptsize(100)} & {\scriptsize(010)} & {\scriptsize(110)} & {\scriptsize(-1-10)} & {\scriptsize(-100)} & {\scriptsize(0-10)} & {\scriptsize(001)} \\
\midrule
$e_g$(1) & $e_g$(1) & $ -77$ & $  0$ & $ -4$ & $  0$ & $  0$ & $  0$ & $ -4$ & $ -4$ \\
$e_g$(2) & $e_g$(2) & $ -77$ & $ -4$ & $  0$ & $  0$ & $  0$ & $ -4$ & $  0$ & $ -4$ \\
$e_g$(2) & $e_g$(1) & $   0$ & $ -4$ & $ -4$ & $  6$ & $  0$ & $  0$ & $  0$ & $  2$ \\
$e_g$(1) & $e_g$(2) & $   0$ & $  0$ & $  0$ & $  0$ & $  6$ & $ -4$ & $ -4$ & $ -2$ \\
$e_g$(1) & O-$p$(1) & $   0$ & $  0$ & $ -3$ & $  0$ & $  0$ & $  3$ & $  0$ & $  0$ \\
$e_g$(1) & O-$p$(2) & $  -7$ & $  0$ & $  0$ & $  0$ & $ -4$ & $  0$ & $  0$ & $  0$ \\
$e_g$(2) & O $p$(1) & $   0$ & $  0$ & $  6$ & $  2$ & $  0$ & $  5$ & $  0$ & $  0$ \\
$e_g$(2) & O $p$(2) & $   0$ & $  0$ & $  3$ & $  2$ & $  0$ & $ -4$ & $  0$ & $  0$ \\
O $p$(1) & O $p$(1) & $-366$ & $ 10$ & $  0$ & $-11$ & $-11$ & $ 10$ & $  0$ & $-85$ \\
O $p$(2) & O $p$(2) & $-366$ & $-10$ & $  0$ & $ 11$ & $ 11$ & $-10$ & $  0$ & $-85$ \\
O $p$(2) & O $p$(1) & $   0$ & $ -4$ & $-22$ & $ 15$ & $ -6$ & $ 18$ & $  0$ & $ -3$ \\
O $p$(1) & O $p$(2) & $   0$ & $ 18$ & $  0$ & $ -6$ & $ 15$ & $ -4$ & $-22$ & $  3$ \\
$a_{1g}$ & $a_{1g}$ & $-715$ & $  6$ & $  6$ & $  6$ & $  6$ & $  6$ & $  6$ & $ 12$ \\
\bottomrule
\end{tabular}

\caption{Five-band model, comprising besides the two Cu $e_g$ orbitals of the two-band model additionally two O $p$(1) and $p$(2) orbitals (corresponding to $p_x$ and $p_y$) as well as one further Cu $a_{1g}$ ($d_{z^2}$) orbital. The numbers in brackets indicate the hopping vector in real space, see Fig.~\ref{Fig3_orbchar}~(f).
Hopping parameters up to 1st nearest neighbor  between all Cu and O sites and orbitals are given. As before,  $t_1$ indicates 1st nearest hoppings, the numbers in brackets indicate hopping vectors in real space; hopping terms obviously smaller than 1\,meV are approximated as 0\,meV. The energy unit is meV.
\label{Tab2_5bands}}
\end{table}

\begin{figure*}[t]
\centering
\includegraphics[width=0.95\textwidth]{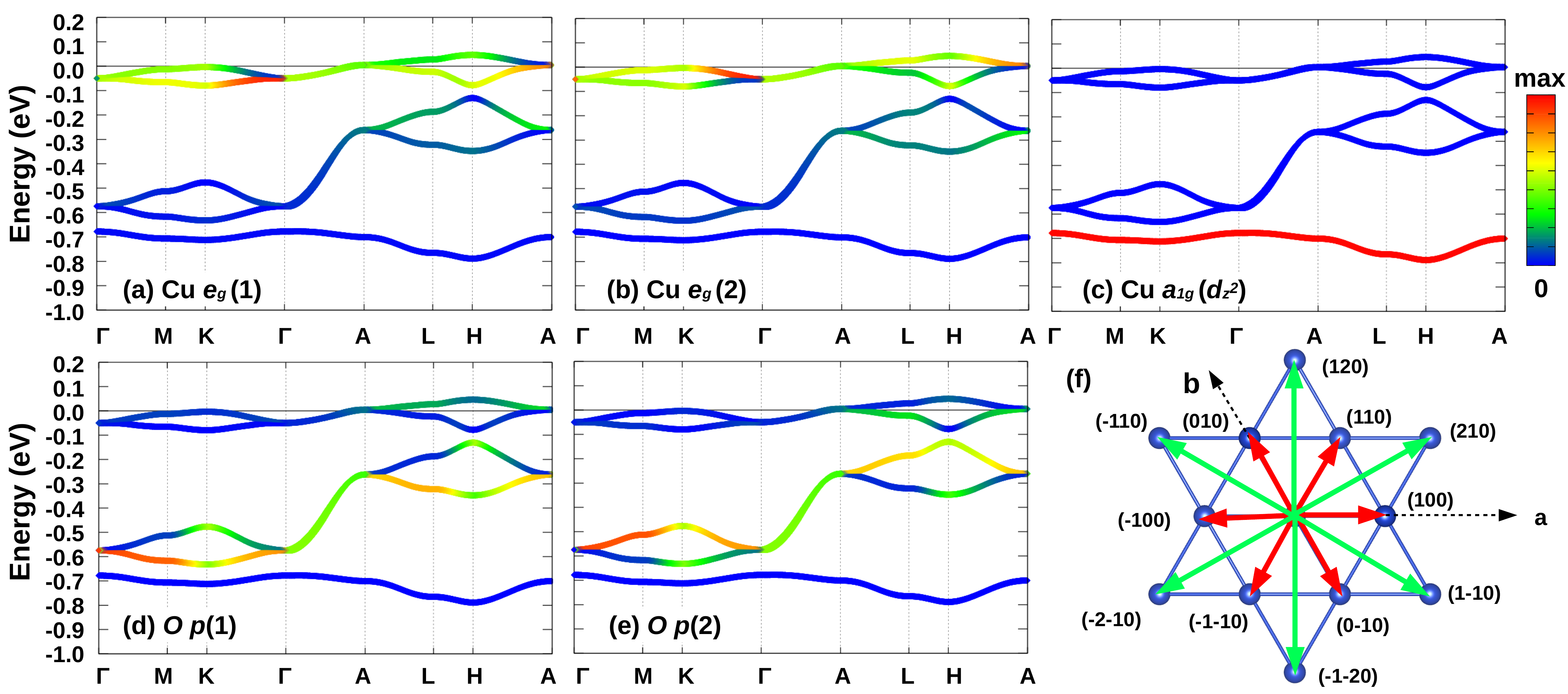}
\caption{Wannier bands character of (a) $e_g$(1), (b) $e_g$(2), (c) $a_{1g}$, (d) O $p$(1), and (e) O $p$(2). (f) Schematic hopping terms of Cu site, the red and green arrows indicate first and second nearest hopping, the numbers in bracket indicate the real space hopping vectors.}
\label{Fig3_orbchar}
\end{figure*}

\begin{figure}[t]
\centering
\includegraphics[width=0.3\textwidth]{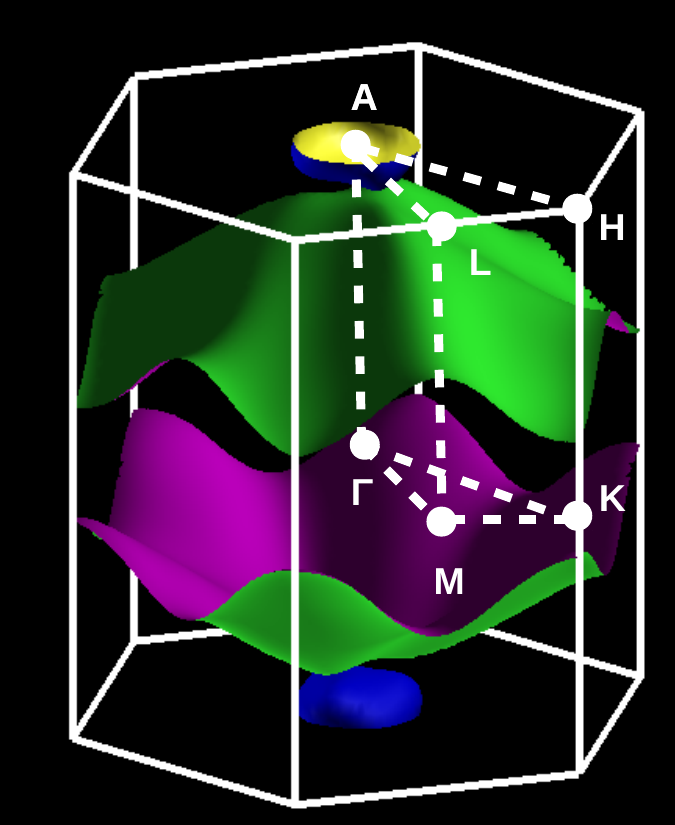}
\caption{(Two-band) tight-binding model  fermi surface of Pb$_9$Cu(PO$_4$)$_6$O.}
\label{Fig4_fermi}
\end{figure}

\section{Computational methods}

DFT-level structural relaxations and static calculations are performed by employing \textsc{Vasp} \cite{PhysRevB.47.558,kresse1996efficiency} and \textsc{Wien2K} \cite{blaha2001wien2k,Schwarz2003} code with the Perdew-Burke-Ernzerhof version for solids of the generalized gradient approximation (GGA-PBESol) \cite{PhysRevLett.100.136406} and a dense 9$\times$9$\times$13 $k$-mesh for the unit cell of Pb$_9$Cu(PO$_4$)$_6$O. The relaxed ground state crystal structure is shown in Fig.~\ref{Fig1_structure}(a). {The two low-energy effective Hamiltonians are} generated by projecting the derived DFT bands, now computed by \textsc{WIEN2K}, around the Fermi level onto Wannier functions \cite{PhysRev.52.191,RevModPhys.84.1419} using \textsc{WIEN2WANNIER}  \cite{mostofi2008wannier90,kunevs2010wien2wannier}. The real-space Wannier Hamiltonian 
is then transformed to momentum space using a $\mathbf{k}$-mesh with 18,125 reducible points.

For the DMFT calculations, this DFT-derived one-particle Hamiltonian is supplemented by a local Kanamori interaction on the Cu sites, see Section \ref{Sec:TB}, and we employ the fully localized limit as double counting correction scheme \cite{Anisimov1991}.
O-$p$ orbitals are considered as non-interacting. We solve the resulting many-body Hamiltonian at room temperature (298\,K, $\beta=1/(k_B T)=39\,\mathrm{eV}^{-1}$) within DMFT employing a continuous-time quantum Monte Carlo (QMC) solver in the hybridization expansions \cite{RevModPhys.83.349} using \textsc{W2dynamics} \cite{PhysRevB.86.155158,wallerberger2019w2dynamics}. Real-frequency spectra are obtained with the \textsc{ana\_cont} code \cite{Kaufmann2021} via analytic continuation using the maximum entropy method (MaxEnt) \cite{PhysRevB.44.6011,PhysRevB.57.10287}.

Further, the rotationally invariant DFT+$U$ scheme~\cite{Liechtenstein_1995} with $U=3$~eV and $J=0.7$~eV 
on top of the regular PBE~\cite{perdew_generalized_1996,perdew_generalized_1997} functional 
as implemented in the \textsc{Vasp} package was
employed for DOS and bandstructure calculations for two representative relaxed crystal structures.
The plane wave cutoff is 600~eV, 4$\times$4$\times$5 and 5$\times$4$\times$4 $k$-mesh 
was used for \emph{P3} (143) and \emph{Pm} (6) structures.
\textsc{AFLOW-SYM}~\cite{Hicks_2018} was used for the symmetry analysis.

\section{Tight binding models}
\label{Sec:TB}
The two-band 
and five-band tight binding  model consist of $m=1..2$ and $m=1..5$ orbitals in the unit cell, respectively. For the two-band model these are the $d_{xz}$ and   $d_{xz}$ orbitals 
of the Cu-site,
for the five-band model there are two additional O $p_x$ and $p_y$ orbitals from
oxygen sites and one more
$d_{z^2}$ orbital from the Cu-site. This {motif} is periodically extended on a hexagonal $k$-grid.  This tight-binding Hamiltonian ${\cal H}^0$ is supplemented by a local Coulomb interaction term ${\cal H}_{int}$ on the Cu sites.

\begin{align}
    {\cal H}= {\cal H}^0+{\cal H}_{int}.
\end{align}
For the non-interacting part 
\begin{align}
    {\cal H}^0=\sum_{\mathbf{k},\sigma,m,n}{\cal H}^0_{m,n}(\mathbf{k}),
\end{align}
we set up a tight-binding parametrization
\begin{align}
    {\cal H}^0_{m,n}(\mathbf{k})= 
    \sum_{i,j} t_{im,jn}e^{i\mathbf{k}(\mathbf{R}_{i}-\mathbf{R}_{j})} c^{\dag}_{i,m\sigma} c_{j,n\sigma}^{\phantom{\dag}}
    ,
\end{align}
where 
$c^{\dag}_{i,m\sigma}(c_{i,m\sigma}^{\phantom{\dag}})$ 
is the creation (annihilation) operator, and $i$, $j$ indicate {unit cells} $\mathbf{R}_{i}$, $\mathbf{R}_{j}$, while $m$, $n$ are orbital indices, and $\sigma$ the spin index. 
For the interaction part, 
\begin{align}
    {\cal H}_{int}=\sum_{i,\sigma,\sigma^\prime} {\cal H}_{int}^{\sigma,\sigma^\prime}(i)
\end{align}
we use the Kanamori form
\begin{eqnarray}
\label{223}
{\cal H}_{int}^{\sigma,\sigma^\prime}(i) &=&U\sum_{m}n_{m,\sigma}n_{m,\sigma'}\\
&+&  \sum_{m>n}[U' n_{m,\sigma}n_{n,\sigma'}+(U'-J)n_{m,\sigma}n_{n,\sigma'}] \nonumber\\
&-&\sum_{m,n}^{m \neq n}J(d^{\dag}_{m,\sigma}d^{\dag}_{n,\sigma'}d_{n,\sigma}d_{m,\sigma'}+d^{\dag}_{m,\sigma}d^{\dag}_{m,\sigma'}d_{n,\sigma}d_{n,\sigma'}),\nonumber
\end{eqnarray}
{where all number operators, $n_{m,\sigma}$, act on the same Cu site (i.e., are in the same unit cell $i$), and the $m,n$ orbitals are restricted to the two and three Cu orbitals for the respective two- and five-band model, defined below.}


\subsection{ Two-band tight-binding model}
As only two bands cross the Fermi energy, our initial objective is to establish a two-band model for Pb$_9$Cu(PO$_4$)$_6$O.
As depicted in Fig.~\ref{Fig1_structure}(a,b), the bands intersecting at the Fermi level (E$_f$) primarily arise from the Cu $d_{yz}$ and $d_{xz}$ orbitals (in the coordinate system with $z$ aligned with the $c$ unit cell vector). Thus, a  {minimal low-energy model with only these two orbitals appears possible. It can facilitate} subsequent calculations that extend beyond the scope of DFT,  as fewer orbitals  {require less} computational resources for complicated many-body calculations. The energy range chosen for {this two-band}  Wannier projections is -0.1\,eV to 0.1\,eV.
Fig.~\ref{Fig1_structure}(c) shows the excellent fit of the bands in the Wannier gauge to the DFT. Truncating the hopping amplitudes at the second nearest neighbors, yields the hopping elements collected in Table~\ref{Tab1_2bands}.

It is worth noting that, due to the presence of a trigonal distortion, the orthogonality between $d_{yz}$ and $d_{xz}$ bands, which is preserved  in an undistorted CuO$_6$ octahedron, is lifted. The  distortion itself can be best seen in Fig.~\ref{Fig1_structure}(b). It leads to the emergence of non-zero hopping terms such as the nearest neighbor ($t_1$) hopping from $e_g$(1) to $e_g$(2) along the (110) direction. For all second nearest neighbor hoppings, the predicted values are close to zero (less than 1\,meV), indicating that hopping between Cu ions in Pb$_9$Cu(PO$_4$)$_6$O can be safely restricted to first nearest neighbors. These nearest Cu neighbors are separated in space by approximately 10\,\AA.

The hopping energy along the $z$-direction, $t_z$, is -10\,meV.
Even this hopping is notably smaller than the corresponding  $t_z$ in infinite-layer nickelates and cuprates ($\sim$-36\,meV; where the in-plane hopping is from -370 to -450\,meV) \cite{Kitatani2020}. 
That is, in contrast to cuprates and nickelates, in-plane and out-of-plane  hoppings are quite comparable;
also the Cu-Cu distance in the $z$-direction is with $\sim$7.4\,\AA~similar to that in-plane ones. 
Altogether, we can hence conclude that  LK-99 has a three dimensional electronic structure.

Furthermore, we illustrate the Fermi surface using the two-band model  in Fig.~\ref{Fig4_fermi}. (The Fermi surface of the five-band model is also exactly the same as that of the original DFT).   Intriguingly, the Fermi surface of LK-99 exhibits striking resemblances to that of UPt$_3$ \cite{de1987upt3,sauls1994order,li2022anomalously}. In UPt$_3$, the prevailing consensus attributes the emergent superconductivity to heavy fermions, rather than electron-phonon coupling. This parallel suggests that the presence of super flat bands and correlations with $U/W$ of the order of ${\cal O}(10)$ might play a pivotal role in driving a transition from a normal to a superconducting state.

\subsection{Five-band tight-binding model}

As the O orbitals may play an important role if LK-99 is doped with electrons or holes, see Fig.~\ref{Fig:sketch}, we further construct a five-band model.
Here, we also include the Cu $d_{z^2}$ orbital to be on the safe side. As long as this lowest band remains firmly below the Fermi surface (i.e., is fully occupied up to hybridizations/orbital admixing), it need not be considered in subsequent many-body calculations.  But it can be included with a simple Hartree shift, given by the occupations (and spin polarizations) of the two Cu $e_g$ orbitals. If on the other hand, this Cu $d_{z^2}$ orbital  accumulates holes, this is indicative that further Cu orbitals besides the two $e_g$ orbitals need to be included in the calculation.

For this extended five-band model, the hopping parameters are detailed in Table~\ref{Tab2_5bands}. 
In congruence with our initial analysis in Fig.\ref{Fig1_structure}(b) and the energy ordering displayed in Fig.\ref{Fig1_structure}(d), the on-site energies ($t_{000}$) 
of Cu $e_g$, O $p$, and Cu $a_{1g}$ ($d_{z^2}$) orbitals amount to -77\,meV, -366\,meV, and -715\,meV, respectively. Notably, focusing on the Cu $a_{1g}$ bands, we observe isotropic hoppings of the order of $\sim$6\,meV along all in-plane directions, while a more substantial hopping is evident along the $z$-direction. This consistently aligns with the inherent symmetry of the $d_{z^2}$ orbital [Fig.\ref{Fig1_structure}(b)]. Furthermore, as depicted in Fig.\ref{Fig3_orbchar}(c), the Cu $a_{1g}$ ($d_{z^2}$) band has a remarkably flat dispersion and exhibits only minimal hybridization with  all other bands, {\em a posteriori} justifying their exclusion from a tight binding model of LK-99.

Turning to the O $p$ orbitals in Table~\ref{Tab2_5bands}, the $p(1)$ to $p(2)$ intra-orbital hoppings can reach up to 11\,meV, while inter-orbital hoppings extend to -22\,meV. A particularly noteworthy observation is the strong intra-orbital hopping between O $p$ orbitals along the $z$-direction ($t_z$), amounting to -85\,meV. This concurs with the pronounced dispersion of O-$p$ orbitals along the $\Gamma$-$A$ path, see Fig.~\ref{Fig1_structure}(d). Considering the distance of approximately $\sim$7.4\,\AA, such a substantial hopping is unexpected and raises the possibility of inducing anisotropic, quasi one-dimensional transport in LK-99 (if it is a charge transfer insulator, and if LK-99 is hole doped).

We can further utilize the tight-binding model for visualizing the band hybridization, by plotting the individual contribution of all five orbitals to the electronic structure. This is shown in Fig.~\ref{Fig3_orbchar}. While the two Cu $e_g$ 
and two O $p$ Wannier orbitals strongly admix among themselves individually, the  mixing between these two types of orbitals as well as between them and  the Cu $d_{z^2}$ orbital is weak. 
A notable exception is the quite strong intermixing (hybridization) between Cu $d_{z^2}$  and O $p$ orbitals around the H k-point.
Despite the very small hopping (hybridization) between these orbitals (e.g., the Cu-O inter-orbital hopping is only $t_1(-100) =3$\,meV in Table~\ref{Tab2_5bands}), the energy difference between the Cu and O bands is similarly small at momentum H. For this reason they still strongly hybridize. For example, in perturbation theory the admixture is  $t_1(-100)/\Delta_H$, where $\Delta_H$ is the energy difference at H.

\subsection{Interaction parameters}

Constrained random phase approximation (cRPA)  \cite{PhysRevB.77.085122} calculations for other
 Cu-$d^9$ based materials \cite{Jang2016,DiSante_cubio,Chang_LSCO,PhysRevB.99.075135} suggest an intra-orbital Hubbard interaction $U=\mathcal{O}(2-3)$\,eV 
 for {the two (three) Cu {$d_{xz/yz}$(+$d_{z^2}$)} orbitals in the 2-band (5-band) calculation.
 Note that the (minimal) Cu-O distance in LK-99 (2.08\AA) is longer by about 10\% compared to typical cuprates, suggesting weaker hybridizations and screening. On the other hand, the bare charge-transfer energy between Cu-$d$ and O-$p$ orbitals, $E_d-E_p$, is much smaller than in cuprates, possibly enhancing screening effects.
 Further, to account for additional spectral-weight transfers from retarded processes, one typically uses a static interaction $U$ that is enhanced with respect to the cRPA-value.
 Therefore, we advocate $U=2.5-3.5$~eV, a Hund's exchange $J=0.7$\,eV, and an inter-orbital interaction $U'=U-2J$. 
 Given the flatness of the relevant copper bands, we expect the parent compound to be insulating for any reasonable interaction strength. The precise value will, however, be important in determining the insulating nature (Mott or charge transfer).

\section{DMFT}

In order to study effects of electronic correlations on top of the tight-binding description, we perform a DMFT  calculation. 
DMFT describes the local dynamics of electrons, by monitoring
the charge (and spin) fluctuations on a given lattice site 
in the presence of the local Hubbard interactions. The latter, in particular, penalizes occupying a site 
with more than one electron per orbital and leads to
a renormalization of the quasi-particle band-structure. 
Or, if the interaction is strong enough, to a Mott-Hubbard splitting of the DFT bands.


Fig.~\ref{fig:aw-0} shows the analytically-continued (to real frequencies) DFT+DMFT spectral function for both (i) the two-band and (ii) the five-band model. 
The (nominal) filling for the models is (i) 
$n=3$ and (ii) $n=9$ electrons per Cu site. 
For both models we consider  two different inter-orbital interactions $U'=2$\,eV and  $U'=3$\,eV, respectively, and  a fixed Hund's exchange  $J=0.7$\,eV; for the intra-orbital (Hubbard) interaction we use $U=U'+2J$. The differences (uncertainties) between $e_g(1)$ and $e_g(2)$ spectra emerge since we did not enforce orbital and spin symmetry.

\begin{figure}
\includegraphics[width=\columnwidth]{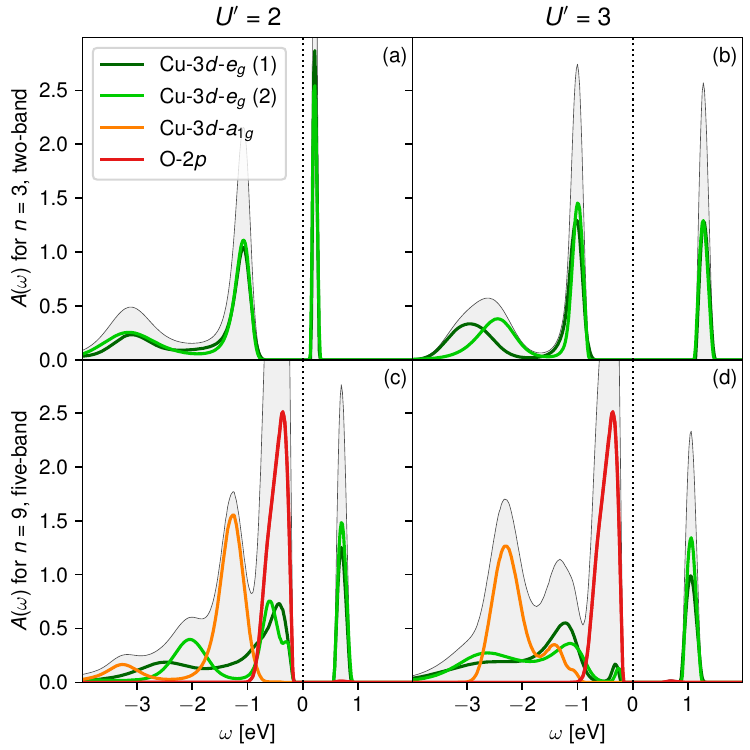}
\caption{Total and orbitally-resolved DMFT spectral function $A(\omega)$ for (a) the  two-band tight-binding model at $U'=2$\,eV and (b) $U'=3$eV, as well as (c) the five-band tight-binding model at $U'=2$\,eV and (d) $U'=3$eV, all at $T=298$K.}
\label{fig:aw-0}
\end{figure}

For both models, we see a clear gap at the Fermi energy. Both  Cu $e_g$ (i.e., $d_{xz}$ and $d_{yz}$) orbitals split into a lower and an upper Hubbard band. Note that despite this Mott-Hubbard splitting, the orbital degeneracy remains. Because the system is not particle-hole symmetric, the weights of the upper and lower Hubbard bands are not symmetric. 

Since the lower Hubbard band describes transitions from 3 electrons on the Cu site to 2 electrons  and since 2 electrons have a singlet-triplet splitting,
the lower Hubbard band has to show a multiplet splitting. Such a splitting is also seen in  Fig.~\ref{fig:aw-0}. In contrast, the transition from 3 electrons to 4 electrons, i.e., the upper Hubbard band, must not show such a splitting. 

The gaps for the two-band model are larger than for the five-band models. 
The reason for this is that in the two-band model three more orbitals contribute to the screening, which reduces the effective interaction in a cRPA calculation. That is, we should, on the very limited $U$ grid available and bare an actual cRPA calculation, rather compare $U'=2$\,eV for the two-band model with $U'=3$\,eV for the five band model.

As for the question of Mott or charge transfer insulator we need to look at the five-band model in Fig.~\ref{fig:aw-0} (c,d):
At $U'=3$\,eV in Fig.~\ref{fig:aw-0} we clearly have a charge transfer insulator, with the first band below the Fermi energy stemming from the oxygen $p$ orbitals. At $U'=2$\,eV we have a very strong
admixture between Cu and O bands in the first spectral contribution below the Fermi energy. The system is neither a clear charge transfer nor a clear Mott insulator: We are right at the crossover between a charge-transfer to a Mott insulator. 
For even smaller $U$'s, given that this tendency continues, the system  might be a Mott insulator. As we do not have interaction parameters from cRPA calculations yet, there is a {\em substantial} uncertainty in $U$; and we cannot give a definite answer whether LK-99 is a Mott or a charge-transfer insulator. As cRPA estimations of $U$ also have a substantial error, it might also be beyond present-day theoretical tools to decide this question.

Let us also note the small peak at  $-0.2$\,eV for $U'=3$\,eV in Fig.~\ref{fig:aw-0} (c). This might be akin to a Zhang-Rice singlet,
see e.g.~\cite{Hansmann2014} for a similar peak in DFT+DMFT calculations for cuprates where such a Zhang-Rice singlet peak is visible. A difference is that,  here, for LK-99, this peak is at the upper edge of the oxygen bands, not above these. Also, we should note that this small peak might be a maximum entropy artifact.

\begin{figure}[tb]
\centering
\includegraphics[width=\columnwidth]{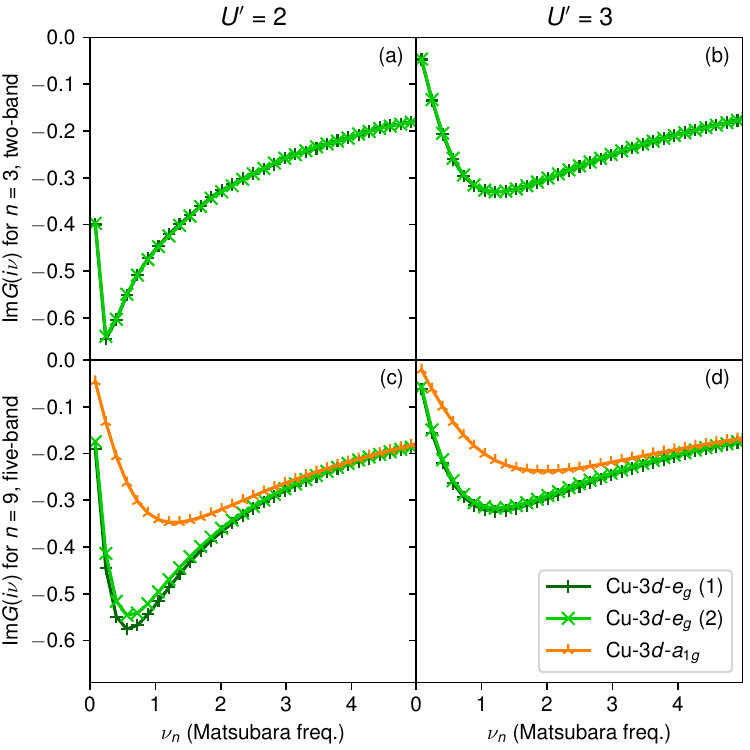}
\caption{Imaginary part of the local DMFT Greens function  $G$ vs.~Matsubara frequency $\nu_n$. Parameters as in Fig.~\ref{fig:aw-0}.}
\label{FigA:G}
\end{figure}

To avoid the maximum entropy uncertainties, we further present in Fig.~\ref{FigA:G}  the   local ($\mathbf k$-integrated)  DMFT Green's function, specifically its imaginary part and now for (imaginary) Matsubara frequencies $\nu_n$. Here $G_{\nu_n}\rightarrow 0$ for $\nu_n\rightarrow 0$ signals that there are no states at the Fermi energy, i.e., we have an insulator. Both, for $n=9$ electrons in the five-band and $n=3$ for the three-band model, $G_{\nu_n}\rightarrow 0$. LK-99 is an insulator.

\section{DFT+$U$}

Previous DFT+$U$ calculations \cite{Lai2023,Kurleto2023} did not find an insulating state for the lattice symmetry and the unit cell also employed here. The degeneracy of the $d_{yz}$ and $d_{xz}$ orbital prevents a splitting since in DFT+$U$ spin or orbital polarization is needed for splitting off Hubbard bands.
Our DFT+$U$ result of Pb$_9$Cu(PO$_4$)$_6$O in Fig.~\ref{fig:dft_u_bands}(a) show the same. 

\begin{figure*}[t]
\centering
\includegraphics[width=1.015\textwidth]{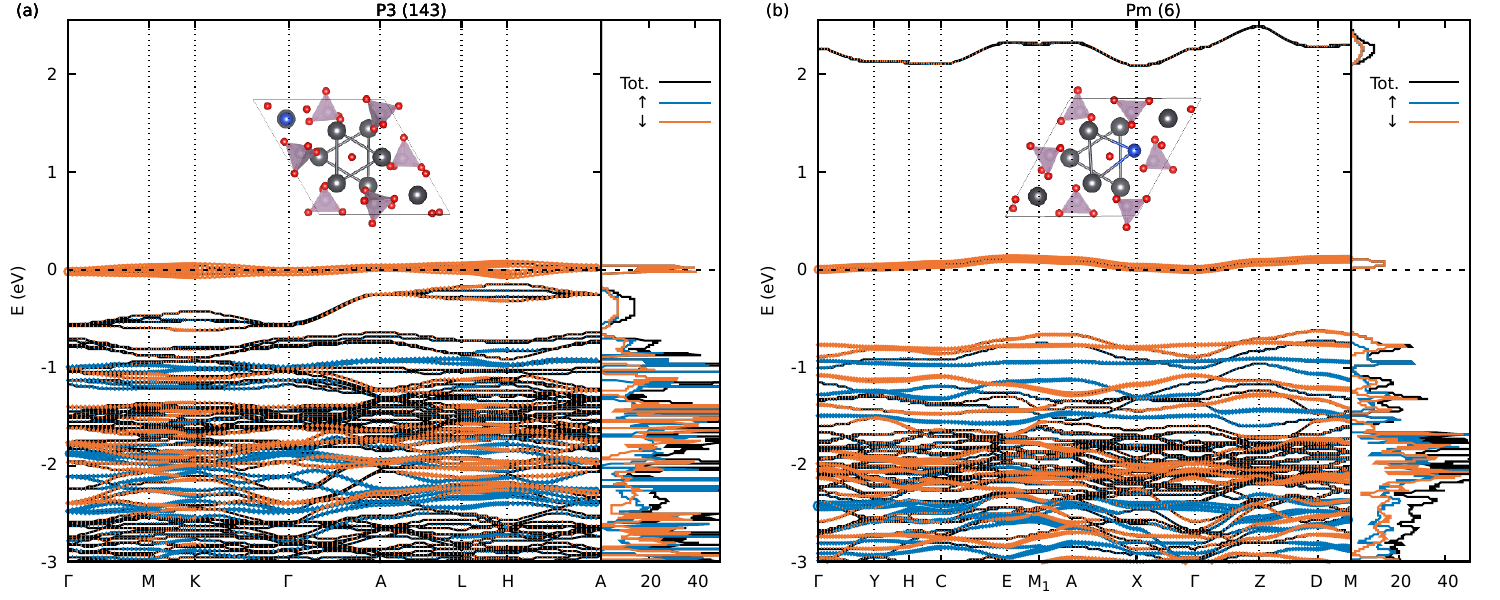}
\caption{DFT+$U$ band structure (left) and DOS (right) for two structures of Pb$_9$Cu(PO$_4$)$_6$O.
Spin-up and spin-down contributions to bands (fatbands) and to DOS are highlighted
in blue and orange.
Panel (a) shows the calculation for the structure presented in Fig.~\ref{Fig1_structure}(a),
while panel (b) is for the structure with a different arrangement of Pb and Cu atoms
(crystal structures are given in the insets with Cu and Pb atoms in blue and gray, respectively).
}
\label{fig:dft_u_bands}
\end{figure*}

However, introducing structural distortion via various positions of Cu-doping, like at Pb(2) in Fig.~\ref{Fig1_structure}(a), further reduced system symmetry from  \emph{P3} (143) of Fig.~\ref{Fig1_structure}(a) to \emph{Pm} (6) after DFT structural relaxations; cf. insets of Fig.~\ref{fig:dft_u_bands}.
In this lower \emph{Pm} symmetry, each Cu atom acquired four surrounding O ligands, leading to a crystal field of strongly distorted tetrahedral shape.
This breaks the orbital degeneracy of the  \emph{P3} (143) structure.  
In DFT+$U$, see Fig.~\ref{fig:dft_u_bands}(b), static electronic correlations then strongly enhance the orbital splitting and additionally yield a spin polarization and a corresponding spin splitting.  This ultimately gives rise to an insulating state with a similar gap as in DMFT, but here with a full spin and orbital polarization. Note that in DFT+$U$ $E_f$ happens to be just below the upper Hubbard band in Fig.~\ref{fig:dft_u_bands}(b).

\section{Conclusion}

To sum up, on the basis of density-functional theory, we performed a Wannier function projection for analyzing the electronic structure of Pb$_9$Cu(PO$_4$)$_6$O (LK-99). This allows for subsequent many-body calculations. Specifically, we constructed two distinct low-energy models: (i) a minimal two-band model including only the Cu $d_{xz}$ and  $d_{yz}$ orbitals and (ii) a five-band model 
that additionally encompasses the O $p_{x}$ and  $p_{y}$ orbitals as well as the Cu $d_{z^2}$ orbital. {These tight binding models shed some } light on LK-99: It has a three-dimensional electronic structure of the two low-energy Cu $d_{xz}$ and $d_{yz}$ bands with next nearest-neighbor hoppings up to distances of 10\,\AA.
This is contrasted with a more one-dimensional bandstructure of the  O $p_x$ and $p_y$ orbitals; the latter orbitals can become relevant if LK-99 is a charge transfer insulator. 
For cuprate superconductors, higher T$_c$'s have been associated with smaller charge-transfer energies  \cite{Weber_2012}, which, there,
are still of the order $\mathcal{O}(1-3)$~eV. 
This empirical trend for cuprates nicely aligns with an allegedly much higher T$_c$ of LK-99, here, since the bare Cu-O  charge-transfer energy in our five-orbital model is an order of magnitude smaller.

The ratio of interaction-to-bandwidth $U/W$ is an order of magnitude larger than what is needed to turn a metal into a Mott insulator \cite{Gebhard1997}. Hence, we see in DMFT a splitting of the Cu $d_{xz}$/$d_{yz}$ into a lower and an upper Hubbard band so that LK-99 becomes either a Mott insulator or charge transfer insulator. Given the uncertainty in $U$ we cannot say, at the moment, which version of these closely related insulators is realized. This splitting  is a dynamical and purely electronic mechanism.  In DMFT we do not observe indications for orbital or spin symmetry breaking. 
In contrast to DMFT, to find an insulator in DFT+$U$, a symmetry breaking of the  Cu $e_g$ ($d_{xz}$/$d_{yz}$) orbitals is required.  For the crystal structure of Fig.~\ref{Fig1_structure}(a) these are, however, degenerate.
In our DFT+$U$ we therefore also considered a structure with Cu on the Pb(1) site in Fig.~\ref{Fig1_structure}(a). In this case, the $e_g$ symmetry is broken and hence the Cu bands can split into Hubbard bands, realizing an insulator also in DFT+$U$.

Such a dynamic splittings into Hubbard bands and dynamical orbital reoccupations 
can match with likewise {\em dynamic} but slower Jahn-Teller phonons  (lattice distortions). For example, in manganites electronic and lattice modes mutually support each other and thus localize charge carriers without symmetry breaking~\cite{Yang2017}. While phonons have not yet been calculated, a similar  or different interplay between electron and phonon dynamics might play an important role for LK-99.

\paragraph*{Acknowledgments}
We thank Robert Svagera and Neven Bari\v{s}i\'{c} for discussions and acknowledge funding through the Austrian Science Fund (FWF) projects I~5398, P~36213, SFB Q-M\&S (FWF project ID F86), the joint project I~6142 of FWF and the French National Research Agency (ANR), and Research Unit QUAST by the Deutsche Foschungsgemeinschaft (DFG; project ID FOR5249) and FWF (project ID I~5868).
L.~S.~is thankful for the starting funds from Northwest University. 
Calculations have been mainly done on the Vienna Scientific Cluster (VSC). 
For the purpose of open access, the authors have applied a CC BY public copyright licence to any Author Accepted Manuscript version arising from this submission.


\begin{thebibliography}{61}%
\makeatletter
\providecommand \@ifxundefined [1]{%
 \@ifx{#1\undefined}
}%
\providecommand \@ifnum [1]{%
 \ifnum #1\expandafter \@firstoftwo
 \else \expandafter \@secondoftwo
 \fi
}%
\providecommand \@ifx [1]{%
 \ifx #1\expandafter \@firstoftwo
 \else \expandafter \@secondoftwo
 \fi
}%
\providecommand \natexlab [1]{#1}%
\providecommand \enquote  [1]{``#1''}%
\providecommand \bibnamefont  [1]{#1}%
\providecommand \bibfnamefont [1]{#1}%
\providecommand \citenamefont [1]{#1}%
\providecommand \href@noop [0]{\@secondoftwo}%
\providecommand \href [0]{\begingroup \@sanitize@url \@href}%
\providecommand \@href[1]{\@@startlink{#1}\@@href}%
\providecommand \@@href[1]{\endgroup#1\@@endlink}%
\providecommand \@sanitize@url [0]{\catcode `\\12\catcode `\$12\catcode
  `\&12\catcode `\#12\catcode `\^12\catcode `\_12\catcode `\%12\relax}%
\providecommand \@@startlink[1]{}%
\providecommand \@@endlink[0]{}%
\providecommand \url  [0]{\begingroup\@sanitize@url \@url }%
\providecommand \@url [1]{\endgroup\@href {#1}{\urlprefix }}%
\providecommand \urlprefix  [0]{URL }%
\providecommand \Eprint [0]{\href }%
\providecommand \doibase [0]{http://dx.doi.org/}%
\providecommand \selectlanguage [0]{\@gobble}%
\providecommand \bibinfo  [0]{\@secondoftwo}%
\providecommand \bibfield  [0]{\@secondoftwo}%
\providecommand \translation [1]{[#1]}%
\providecommand \BibitemOpen [0]{}%
\providecommand \bibitemStop [0]{}%
\providecommand \bibitemNoStop [0]{.\EOS\space}%
\providecommand \EOS [0]{\spacefactor3000\relax}%
\providecommand \BibitemShut  [1]{\csname bibitem#1\endcsname}%
\let\auto@bib@innerbib\@empty
\bibitem [{\citenamefont {{Lee}}\ \emph
  {et~al.}(2023{\natexlab{a}})\citenamefont {{Lee}}, \citenamefont {{Kim}},\
  and\ \citenamefont {{Kwon}}}]{Lee2023_2}%
  \BibitemOpen
  \bibfield  {author} {\bibinfo {author} {\bibfnamefont {S.}~\bibnamefont
  {{Lee}}}, \bibinfo {author} {\bibfnamefont {J.-H.}\ \bibnamefont {{Kim}}}, \
  and\ \bibinfo {author} {\bibfnamefont {Y.-W.}\ \bibnamefont {{Kwon}}},\
  }\href {https://doi.org/10.48550/arXiv.2307.12008} {\bibfield  {journal}
  {\bibinfo  {journal} {arXiv:2307.12008}\ } (\bibinfo {year}
  {2023}{\natexlab{a}})}\BibitemShut {NoStop}%
\bibitem [{\citenamefont {{Lee}}\ \emph
  {et~al.}(2023{\natexlab{b}})\citenamefont {{Lee}}, \citenamefont {{Kim}},
  \citenamefont {{Kim}}, \citenamefont {{Im}}, \citenamefont {{An}},\ and\
  \citenamefont {{Auh}}}]{Lee2023_3}%
  \BibitemOpen
  \bibfield  {author} {\bibinfo {author} {\bibfnamefont {S.}~\bibnamefont
  {{Lee}}}, \bibinfo {author} {\bibfnamefont {J.}~\bibnamefont {{Kim}}},
  \bibinfo {author} {\bibfnamefont {H.-T.}\ \bibnamefont {{Kim}}}, \bibinfo
  {author} {\bibfnamefont {S.}~\bibnamefont {{Im}}}, \bibinfo {author}
  {\bibfnamefont {S.}~\bibnamefont {{An}}}, \ and\ \bibinfo {author}
  {\bibfnamefont {K.~H.}\ \bibnamefont {{Auh}}},\ }\href
  {https://doi.org/10.48550/arXiv.2307.12037} {\bibfield  {journal} {\bibinfo
  {journal} {arXiv:2307.12037}\ } (\bibinfo {year}
  {2023}{\natexlab{b}})}\BibitemShut {NoStop}%
\bibitem [{\citenamefont {{Lee}}\ \emph
  {et~al.}(2023{\natexlab{c}})\citenamefont {{Lee}}, \citenamefont {{Kim}},
  \citenamefont {{Im}}, \citenamefont {{An}}, \citenamefont {Kwon},\ and\
  \citenamefont {{Auh}}}]{Lee2023_1}%
  \BibitemOpen
  \bibfield  {author} {\bibinfo {author} {\bibfnamefont {S.}~\bibnamefont
  {{Lee}}}, \bibinfo {author} {\bibfnamefont {J.}~\bibnamefont {{Kim}}},
  \bibinfo {author} {\bibfnamefont {S.}~\bibnamefont {{Im}}}, \bibinfo {author}
  {\bibfnamefont {S.}~\bibnamefont {{An}}}, \bibinfo {author} {\bibfnamefont
  {Y.-W.}\ \bibnamefont {Kwon}}, \ and\ \bibinfo {author} {\bibfnamefont
  {K.~H.}\ \bibnamefont {{Auh}}},\ }\href {\doibase
  10.6111/JKCGCT.2023.33.2.061} {\bibfield  {journal} {\bibinfo  {journal} {J.
  Korean Cryst. Growth Cryst. Technol.}\ }\textbf {\bibinfo {volume} {33}},\
  \bibinfo {pages} {61} (\bibinfo {year} {2023}{\natexlab{c}})}\BibitemShut
  {NoStop}%
\bibitem [{\citenamefont {{Wu}}\ \emph {et~al.}(2023)\citenamefont {{Wu}},
  \citenamefont {{Yang}}, \citenamefont {{Xiao}},\ and\ \citenamefont
  {{Chang}}}]{Wu2023}%
  \BibitemOpen
  \bibfield  {author} {\bibinfo {author} {\bibfnamefont {H.}~\bibnamefont
  {{Wu}}}, \bibinfo {author} {\bibfnamefont {L.}~\bibnamefont {{Yang}}},
  \bibinfo {author} {\bibfnamefont {B.}~\bibnamefont {{Xiao}}}, \ and\ \bibinfo
  {author} {\bibfnamefont {H.}~\bibnamefont {{Chang}}},\ }\href
  {https://doi.org/10.48550/arXiv.2308.01516} {\bibfield  {journal} {\bibinfo
  {journal} {arXiv:2308.01516}\ } (\bibinfo {year} {2023})}\BibitemShut
  {NoStop}%
\bibitem [{\citenamefont {{Hou}}\ \emph {et~al.}(2023)\citenamefont {{Hou}},
  \citenamefont {{Wei}}, \citenamefont {{Zhou}}, \citenamefont {{Sun}},\ and\
  \citenamefont {{Shi}}}]{Hou2023}%
  \BibitemOpen
  \bibfield  {author} {\bibinfo {author} {\bibfnamefont {Q.}~\bibnamefont
  {{Hou}}}, \bibinfo {author} {\bibfnamefont {W.}~\bibnamefont {{Wei}}},
  \bibinfo {author} {\bibfnamefont {X.}~\bibnamefont {{Zhou}}}, \bibinfo
  {author} {\bibfnamefont {Y.}~\bibnamefont {{Sun}}}, \ and\ \bibinfo {author}
  {\bibfnamefont {Z.}~\bibnamefont {{Shi}}},\ }\href
  {https://doi.org/10.48550/arXiv.2308.01192} {\bibfield  {journal} {\bibinfo
  {journal} {arXiv:2308.01192}\ } (\bibinfo {year} {2023})}\BibitemShut
  {NoStop}%
\bibitem [{\citenamefont {{Liu}}\ \emph {et~al.}(2023)\citenamefont {{Liu}},
  \citenamefont {{Meng}}, \citenamefont {{Wang}}, \citenamefont {{Chen}},
  \citenamefont {{Duan}}, \citenamefont {{Zhou}}, \citenamefont {{Yan}},
  \citenamefont {{Qin}},\ and\ \citenamefont {{Liu}}}]{Liu2023}%
  \BibitemOpen
  \bibfield  {author} {\bibinfo {author} {\bibfnamefont {L.}~\bibnamefont
  {{Liu}}}, \bibinfo {author} {\bibfnamefont {Z.}~\bibnamefont {{Meng}}},
  \bibinfo {author} {\bibfnamefont {X.}~\bibnamefont {{Wang}}}, \bibinfo
  {author} {\bibfnamefont {H.}~\bibnamefont {{Chen}}}, \bibinfo {author}
  {\bibfnamefont {Z.}~\bibnamefont {{Duan}}}, \bibinfo {author} {\bibfnamefont
  {X.}~\bibnamefont {{Zhou}}}, \bibinfo {author} {\bibfnamefont
  {H.}~\bibnamefont {{Yan}}}, \bibinfo {author} {\bibfnamefont
  {P.}~\bibnamefont {{Qin}}}, \ and\ \bibinfo {author} {\bibfnamefont
  {Z.}~\bibnamefont {{Liu}}},\ }\href
  {https://doi.org/10.48550/arXiv.2307.16802} {\bibfield  {journal} {\bibinfo
  {journal} {arXiv:2307.16802}\ } (\bibinfo {year} {2023})}\BibitemShut
  {NoStop}%
\bibitem [{\citenamefont {{Kumar}}\ \emph {et~al.}(2023)\citenamefont
  {{Kumar}}, \citenamefont {{Karn}},\ and\ \citenamefont
  {{Awana}}}]{Kumar2023}%
  \BibitemOpen
  \bibfield  {author} {\bibinfo {author} {\bibfnamefont {K.}~\bibnamefont
  {{Kumar}}}, \bibinfo {author} {\bibfnamefont {N.~K.}\ \bibnamefont {{Karn}}},
  \ and\ \bibinfo {author} {\bibfnamefont {V.~P.~S.}\ \bibnamefont {{Awana}}},\
  }\href {https://doi.org/10.48550/arXiv.2307.16402} {\bibfield  {journal}
  {\bibinfo  {journal} {arXiv:2307.16402}\ } (\bibinfo {year}
  {2023})}\BibitemShut {NoStop}%
\bibitem [{\citenamefont {Kumar}\ \emph {et~al.}(2023)\citenamefont {Kumar},
  \citenamefont {Karn}, \citenamefont {Kumar},\ and\ \citenamefont
  {Awana}}]{Kumar2023_2}%
  \BibitemOpen
  \bibfield  {author} {\bibinfo {author} {\bibfnamefont {K.}~\bibnamefont
  {Kumar}}, \bibinfo {author} {\bibfnamefont {N.~K.}\ \bibnamefont {Karn}},
  \bibinfo {author} {\bibfnamefont {Y.}~\bibnamefont {Kumar}}, \ and\ \bibinfo
  {author} {\bibfnamefont {V.~P.~S.}\ \bibnamefont {Awana}},\ }\href
  {https://doi.org/10.48550/arXiv.2308.03544} {\bibfield  {journal} {\bibinfo
  {journal} {arXiv:2308.03544}\ } (\bibinfo {year} {2023})}\BibitemShut
  {NoStop}%
\bibitem [{\citenamefont {{Abramian}}\ \emph {et~al.}(2023)\citenamefont
  {{Abramian}}, \citenamefont {{Kuzanyan}}, \citenamefont {{Nikoghosyan}},
  \citenamefont {{Teknowijoyo}},\ and\ \citenamefont
  {{Gulian}}}]{Abramian2023}%
  \BibitemOpen
  \bibfield  {author} {\bibinfo {author} {\bibfnamefont {P.}~\bibnamefont
  {{Abramian}}}, \bibinfo {author} {\bibfnamefont {A.}~\bibnamefont
  {{Kuzanyan}}}, \bibinfo {author} {\bibfnamefont {V.}~\bibnamefont
  {{Nikoghosyan}}}, \bibinfo {author} {\bibfnamefont {S.}~\bibnamefont
  {{Teknowijoyo}}}, \ and\ \bibinfo {author} {\bibfnamefont {A.}~\bibnamefont
  {{Gulian}}},\ }\href {https://doi.org/10.48550/arXiv.2308.01723} {\bibfield
  {journal} {\bibinfo  {journal} {arXiv:2308.01723}\ } (\bibinfo {year}
  {2023})}\BibitemShut {NoStop}%
\bibitem [{\citenamefont {Guo}\ \emph {et~al.}(2023)\citenamefont {Guo},
  \citenamefont {Li},\ and\ \citenamefont {Jia}}]{Guo2023}%
  \BibitemOpen
  \bibfield  {author} {\bibinfo {author} {\bibfnamefont {K.}~\bibnamefont
  {Guo}}, \bibinfo {author} {\bibfnamefont {Y.}~\bibnamefont {Li}}, \ and\
  \bibinfo {author} {\bibfnamefont {S.}~\bibnamefont {Jia}},\ }\href
  {https://doi.org/10.48550/arXiv.2308.03110} {\bibfield  {journal} {\bibinfo
  {journal} {arXiv:2308.03110}\ } (\bibinfo {year} {2023})}\BibitemShut
  {NoStop}%
\bibitem [{\citenamefont {Hohenberg}\ and\ \citenamefont
  {Kohn}(1964)}]{PhysRev.136.B864}%
  \BibitemOpen
  \bibfield  {author} {\bibinfo {author} {\bibfnamefont {P.}~\bibnamefont
  {Hohenberg}}\ and\ \bibinfo {author} {\bibfnamefont {W.}~\bibnamefont
  {Kohn}},\ }\href {\doibase 10.1103/PhysRev.136.B864} {\bibfield  {journal}
  {\bibinfo  {journal} {Phys. Rev.}\ }\textbf {\bibinfo {volume} {136}},\
  \bibinfo {pages} {B864} (\bibinfo {year} {1964})}\BibitemShut {NoStop}%
\bibitem [{\citenamefont {{Lai}}\ \emph {et~al.}(2023)\citenamefont {{Lai}},
  \citenamefont {{Li}}, \citenamefont {{Liu}}, \citenamefont {{Sun}},\ and\
  \citenamefont {{Chen}}}]{Lai2023}%
  \BibitemOpen
  \bibfield  {author} {\bibinfo {author} {\bibfnamefont {J.}~\bibnamefont
  {{Lai}}}, \bibinfo {author} {\bibfnamefont {J.}~\bibnamefont {{Li}}},
  \bibinfo {author} {\bibfnamefont {P.}~\bibnamefont {{Liu}}}, \bibinfo
  {author} {\bibfnamefont {Y.}~\bibnamefont {{Sun}}}, \ and\ \bibinfo {author}
  {\bibfnamefont {X.-Q.}\ \bibnamefont {{Chen}}},\ }\href
  {https://doi.org/10.48550/arXiv.2307.16040} {\bibfield  {journal} {\bibinfo
  {journal} {arXiv:2307.16040}\ } (\bibinfo {year} {2023})}\BibitemShut
  {NoStop}%
\bibitem [{\citenamefont {{Griffin}}(2023)}]{Griffin2023}%
  \BibitemOpen
  \bibfield  {author} {\bibinfo {author} {\bibfnamefont {S.~M.}\ \bibnamefont
  {{Griffin}}},\ }\href {https://doi.org/10.48550/arXiv.2307.16892} {\bibfield
  {journal} {\bibinfo  {journal} {arXiv:2307.16892}\ } (\bibinfo {year}
  {2023})}\BibitemShut {NoStop}%
\bibitem [{\citenamefont {{Si}}\ and\ \citenamefont {{Held}}(2023)}]{Si2023}%
  \BibitemOpen
  \bibfield  {author} {\bibinfo {author} {\bibfnamefont {L.}~\bibnamefont
  {{Si}}}\ and\ \bibinfo {author} {\bibfnamefont {K.}~\bibnamefont {{Held}}},\
  }\href {https://doi.org/10.48550/arXiv.2308.00676} {\bibfield  {journal}
  {\bibinfo  {journal} {arXiv:2308.00676}\ } (\bibinfo {year}
  {2023})}\BibitemShut {NoStop}%
\bibitem [{\citenamefont {{Kurleto}}\ \emph {et~al.}(2023)\citenamefont
  {{Kurleto}}, \citenamefont {{Lany}}, \citenamefont {{Pashov}}, \citenamefont
  {{Acharya}}, \citenamefont {{van Schilfgaarde}},\ and\ \citenamefont
  {{Dessau}}}]{Kurleto2023}%
  \BibitemOpen
  \bibfield  {author} {\bibinfo {author} {\bibfnamefont {R.}~\bibnamefont
  {{Kurleto}}}, \bibinfo {author} {\bibfnamefont {S.}~\bibnamefont {{Lany}}},
  \bibinfo {author} {\bibfnamefont {D.}~\bibnamefont {{Pashov}}}, \bibinfo
  {author} {\bibfnamefont {S.}~\bibnamefont {{Acharya}}}, \bibinfo {author}
  {\bibfnamefont {M.}~\bibnamefont {{van Schilfgaarde}}}, \ and\ \bibinfo
  {author} {\bibfnamefont {D.~S.}\ \bibnamefont {{Dessau}}},\ }\href
  {https://doi.org/10.48550/arXiv.2308.00698} {\bibfield  {journal} {\bibinfo
  {journal} {arXiv:2308.00698}\ } (\bibinfo {year} {2023})}\BibitemShut
  {NoStop}%
\bibitem [{\citenamefont {{Cabezas-Escares}}\ \emph {et~al.}(2023)\citenamefont
  {{Cabezas-Escares}}, \citenamefont {{Barrera}}, \citenamefont {{Cardenas}},\
  and\ \citenamefont {{Munoz}}}]{CabezasEscares2023}%
  \BibitemOpen
  \bibfield  {author} {\bibinfo {author} {\bibfnamefont {J.}~\bibnamefont
  {{Cabezas-Escares}}}, \bibinfo {author} {\bibfnamefont {N.~F.}\ \bibnamefont
  {{Barrera}}}, \bibinfo {author} {\bibfnamefont {C.}~\bibnamefont
  {{Cardenas}}}, \ and\ \bibinfo {author} {\bibfnamefont {F.}~\bibnamefont
  {{Munoz}}},\ }\href {https://doi.org/10.48550/arXiv.2308.01135} {\bibfield
  {journal} {\bibinfo  {journal} {arXiv:2308.01135}\ } (\bibinfo {year}
  {2023})}\BibitemShut {NoStop}%
\bibitem [{\citenamefont {Tao}\ \emph {et~al.}(2023)\citenamefont {Tao},
  \citenamefont {Chen}, \citenamefont {Yang}, \citenamefont {Gao},
  \citenamefont {Xue},\ and\ \citenamefont {Jia}}]{Tao2023}%
  \BibitemOpen
  \bibfield  {author} {\bibinfo {author} {\bibfnamefont {K.}~\bibnamefont
  {Tao}}, \bibinfo {author} {\bibfnamefont {R.}~\bibnamefont {Chen}}, \bibinfo
  {author} {\bibfnamefont {L.}~\bibnamefont {Yang}}, \bibinfo {author}
  {\bibfnamefont {J.}~\bibnamefont {Gao}}, \bibinfo {author} {\bibfnamefont
  {D.}~\bibnamefont {Xue}}, \ and\ \bibinfo {author} {\bibfnamefont
  {C.}~\bibnamefont {Jia}},\ }\href {https://doi.org/10.48550/arXiv.2308.03218}
  {\bibfield  {journal} {\bibinfo  {journal} {arXiv:2308.03218}\ } (\bibinfo
  {year} {2023})}\BibitemShut {NoStop}%
\bibitem [{\citenamefont {Sun}\ \emph {et~al.}(2023)\citenamefont {Sun},
  \citenamefont {Ho},\ and\ \citenamefont {Antropov}}]{Sun2023}%
  \BibitemOpen
  \bibfield  {author} {\bibinfo {author} {\bibfnamefont {Y.}~\bibnamefont
  {Sun}}, \bibinfo {author} {\bibfnamefont {K.-M.}\ \bibnamefont {Ho}}, \ and\
  \bibinfo {author} {\bibfnamefont {V.}~\bibnamefont {Antropov}},\ }\href
  {https://doi.org/10.48550/arXiv.2308.03454} {\bibfield  {journal} {\bibinfo
  {journal} {arXiv:2308.03454}\ } (\bibinfo {year} {2023})}\BibitemShut
  {NoStop}%
\bibitem [{\citenamefont {Kuroki}\ \emph {et~al.}(2005)\citenamefont {Kuroki},
  \citenamefont {Higashida},\ and\ \citenamefont {Arita}}]{Kuroki2005}%
  \BibitemOpen
  \bibfield  {author} {\bibinfo {author} {\bibfnamefont {K.}~\bibnamefont
  {Kuroki}}, \bibinfo {author} {\bibfnamefont {T.}~\bibnamefont {Higashida}}, \
  and\ \bibinfo {author} {\bibfnamefont {R.}~\bibnamefont {Arita}},\ }\href
  {\doibase 10.1103/PhysRevB.72.212509} {\bibfield  {journal} {\bibinfo
  {journal} {Phys. Rev. B}\ }\textbf {\bibinfo {volume} {72}},\ \bibinfo
  {pages} {212509} (\bibinfo {year} {2005})}\BibitemShut {NoStop}%
\bibitem [{\citenamefont {Iglovikov}\ \emph {et~al.}(2014)\citenamefont
  {Iglovikov}, \citenamefont {H\'ebert}, \citenamefont {Gr\'emaud},
  \citenamefont {Batrouni},\ and\ \citenamefont {Scalettar}}]{Iglovikov2014}%
  \BibitemOpen
  \bibfield  {author} {\bibinfo {author} {\bibfnamefont {V.~I.}\ \bibnamefont
  {Iglovikov}}, \bibinfo {author} {\bibfnamefont {F.}~\bibnamefont {H\'ebert}},
  \bibinfo {author} {\bibfnamefont {B.}~\bibnamefont {Gr\'emaud}}, \bibinfo
  {author} {\bibfnamefont {G.~G.}\ \bibnamefont {Batrouni}}, \ and\ \bibinfo
  {author} {\bibfnamefont {R.~T.}\ \bibnamefont {Scalettar}},\ }\href {\doibase
  10.1103/PhysRevB.90.094506} {\bibfield  {journal} {\bibinfo  {journal} {Phys.
  Rev. B}\ }\textbf {\bibinfo {volume} {90}},\ \bibinfo {pages} {094506}
  (\bibinfo {year} {2014})}\BibitemShut {NoStop}%
\bibitem [{\citenamefont {Aoki}(2020)}]{Aoki2020}%
  \BibitemOpen
  \bibfield  {author} {\bibinfo {author} {\bibfnamefont {H.}~\bibnamefont
  {Aoki}},\ }\href {\doibase 10.1007/s10948-020-05474-6} {\bibfield  {journal}
  {\bibinfo  {journal} {J. Supercond. Nov. Magn.}\ }\textbf {\bibinfo {volume}
  {33}},\ \bibinfo {pages} {2341} (\bibinfo {year} {2020})}\BibitemShut
  {NoStop}%
\bibitem [{\citenamefont {Gebhard}(1997)}]{Gebhard1997}%
  \BibitemOpen
  \bibfield  {author} {\bibinfo {author} {\bibfnamefont {F.}~\bibnamefont
  {Gebhard}},\ }\href {https://books.google.at/books?id=HyzyBwAAQBAJ} {\emph
  {\bibinfo {title} {The Mott Metal-insulator transition}}}\ (\bibinfo
  {publisher} {Springer-Verlag (Berlin)},\ \bibinfo {year} {1997})\BibitemShut
  {NoStop}%
\bibitem [{\citenamefont {Zaanen}\ \emph {et~al.}(1985)\citenamefont {Zaanen},
  \citenamefont {Sawatzky},\ and\ \citenamefont {Allen}}]{Zaanen1985}%
  \BibitemOpen
  \bibfield  {author} {\bibinfo {author} {\bibfnamefont {J.}~\bibnamefont
  {Zaanen}}, \bibinfo {author} {\bibfnamefont {G.~A.}\ \bibnamefont
  {Sawatzky}}, \ and\ \bibinfo {author} {\bibfnamefont {J.~W.}\ \bibnamefont
  {Allen}},\ }\href {\doibase 10.1103/PhysRevLett.55.418} {\bibfield  {journal}
  {\bibinfo  {journal} {Phys. Rev. Lett.}\ }\textbf {\bibinfo {volume} {55}},\
  \bibinfo {pages} {418} (\bibinfo {year} {1985})}\BibitemShut {NoStop}%
\bibitem [{\citenamefont {Bednorz}\ and\ \citenamefont
  {M\"uller}(1986)}]{Bednorz1986}%
  \BibitemOpen
  \bibfield  {author} {\bibinfo {author} {\bibfnamefont {J.~G.}\ \bibnamefont
  {Bednorz}}\ and\ \bibinfo {author} {\bibfnamefont {K.~A.}\ \bibnamefont
  {M\"uller}},\ }\href {https://link.springer.com/article/10.1007%2FBF01303701}
  {\bibfield  {journal} {\bibinfo  {journal} {Zeitschrift f\"ur Physik B
  Condensed Matter}\ }\textbf {\bibinfo {volume} {64}},\ \bibinfo {pages} {189}
  (\bibinfo {year} {1986})}\BibitemShut {NoStop}%
\bibitem [{\citenamefont {Scalapino}(2012)}]{Scalapino2012}%
  \BibitemOpen
  \bibfield  {author} {\bibinfo {author} {\bibfnamefont {D.~J.}\ \bibnamefont
  {Scalapino}},\ }\href {\doibase 10.1103/RevModPhys.84.1383} {\bibfield
  {journal} {\bibinfo  {journal} {Rev. Mod. Phys.}\ }\textbf {\bibinfo {volume}
  {84}},\ \bibinfo {pages} {1383} (\bibinfo {year} {2012})}\BibitemShut
  {NoStop}%
\bibitem [{\citenamefont {{Baskaran}}(2023)}]{Baskaran2023}%
  \BibitemOpen
  \bibfield  {author} {\bibinfo {author} {\bibfnamefont {G.}~\bibnamefont
  {{Baskaran}}},\ }\href {https://doi.org/10.48550/arXiv.2308.01307} {\bibfield
   {journal} {\bibinfo  {journal} {arXiv:2308.01307}\ } (\bibinfo {year}
  {2023})}\BibitemShut {NoStop}%
\bibitem [{\citenamefont {Bardeen}\ \emph {et~al.}(1957)\citenamefont
  {Bardeen}, \citenamefont {Cooper},\ and\ \citenamefont
  {Schrieffer}}]{BCS1957}%
  \BibitemOpen
  \bibfield  {author} {\bibinfo {author} {\bibfnamefont {J.}~\bibnamefont
  {Bardeen}}, \bibinfo {author} {\bibfnamefont {L.~N.}\ \bibnamefont {Cooper}},
  \ and\ \bibinfo {author} {\bibfnamefont {J.~R.}\ \bibnamefont {Schrieffer}},\
  }\href {\doibase 10.1103/PhysRev.106.162} {\bibfield  {journal} {\bibinfo
  {journal} {Phys. Rev.}\ }\textbf {\bibinfo {volume} {106}},\ \bibinfo {pages}
  {162} (\bibinfo {year} {1957})}\BibitemShut {NoStop}%
\bibitem [{\citenamefont {{Tavakol}}\ and\ \citenamefont
  {{Scaffidi}}(2023)}]{Tavakol2023}%
  \BibitemOpen
  \bibfield  {author} {\bibinfo {author} {\bibfnamefont {O.}~\bibnamefont
  {{Tavakol}}}\ and\ \bibinfo {author} {\bibfnamefont {T.}~\bibnamefont
  {{Scaffidi}}},\ }\href {https://doi.org/10.48550/arXiv.2308.01315} {\bibfield
   {journal} {\bibinfo  {journal} {arXiv:2308.01315}\ } (\bibinfo {year}
  {2023})}\BibitemShut {NoStop}%
\bibitem [{\citenamefont {Oh}\ and\ \citenamefont {Zhang}(2023)}]{Oh2023}%
  \BibitemOpen
  \bibfield  {author} {\bibinfo {author} {\bibfnamefont {H.}~\bibnamefont
  {Oh}}\ and\ \bibinfo {author} {\bibfnamefont {Y.-H.}\ \bibnamefont {Zhang}},\
  }\href {https://doi.org/10.48550/arXiv.2308.02469} {\bibfield  {journal}
  {\bibinfo  {journal} {arXiv:2308.02469}\ } (\bibinfo {year}
  {2023})}\BibitemShut {NoStop}%
\bibitem [{\citenamefont {Kresse}\ and\ \citenamefont
  {Hafner}(1993)}]{PhysRevB.47.558}%
  \BibitemOpen
  \bibfield  {author} {\bibinfo {author} {\bibfnamefont {G.}~\bibnamefont
  {Kresse}}\ and\ \bibinfo {author} {\bibfnamefont {J.}~\bibnamefont
  {Hafner}},\ }\href {\doibase 10.1103/PhysRevB.47.558} {\bibfield  {journal}
  {\bibinfo  {journal} {Phys. Rev. B}\ }\textbf {\bibinfo {volume} {47}},\
  \bibinfo {pages} {558} (\bibinfo {year} {1993})}\BibitemShut {NoStop}%
\bibitem [{\citenamefont {Kresse}\ and\ \citenamefont
  {Furthm{\"u}ller}(1996)}]{kresse1996efficiency}%
  \BibitemOpen
  \bibfield  {author} {\bibinfo {author} {\bibfnamefont {G.}~\bibnamefont
  {Kresse}}\ and\ \bibinfo {author} {\bibfnamefont {J.}~\bibnamefont
  {Furthm{\"u}ller}},\ }\href {\doibase 10.1016/0927-0256(96)00008-0}
  {\bibfield  {journal} {\bibinfo  {journal} {Computational materials science}\
  }\textbf {\bibinfo {volume} {6}},\ \bibinfo {pages} {15} (\bibinfo {year}
  {1996})}\BibitemShut {NoStop}%
\bibitem [{\citenamefont {Blaha}\ \emph {et~al.}(2001)\citenamefont {Blaha},
  \citenamefont {Schwarz}, \citenamefont {Madsen}, \citenamefont {Kvasnicka},\
  and\ \citenamefont {Luitz}}]{blaha2001wien2k}%
  \BibitemOpen
  \bibfield  {author} {\bibinfo {author} {\bibfnamefont {P.}~\bibnamefont
  {Blaha}}, \bibinfo {author} {\bibfnamefont {K.}~\bibnamefont {Schwarz}},
  \bibinfo {author} {\bibfnamefont {G.}~\bibnamefont {Madsen}}, \bibinfo
  {author} {\bibfnamefont {D.}~\bibnamefont {Kvasnicka}}, \ and\ \bibinfo
  {author} {\bibfnamefont {J.}~\bibnamefont {Luitz}},\ }\href@noop {}
  {\bibfield  {journal} {\bibinfo  {journal} {An augmented plane wave + local
  orbitals program for calculating crystal properties}\ } (\bibinfo {year}
  {2001})}\BibitemShut {NoStop}%
\bibitem [{\citenamefont {Schwarz}\ and\ \citenamefont
  {Blaha}(2003)}]{Schwarz2003}%
  \BibitemOpen
  \bibfield  {author} {\bibinfo {author} {\bibfnamefont {K.}~\bibnamefont
  {Schwarz}}\ and\ \bibinfo {author} {\bibfnamefont {P.}~\bibnamefont
  {Blaha}},\ }\href {\doibase https://doi.org/10.1016/S0927-0256(03)00112-5}
  {\bibfield  {journal} {\bibinfo  {journal} {Computational Materials Science}\
  }\textbf {\bibinfo {volume} {28}},\ \bibinfo {pages} {259} (\bibinfo {year}
  {2003})},\ \bibinfo {note} {{P}roceedings of the {S}ymposium on {S}oftware
  {D}evelopment for {P}rocess and {M}aterials {D}esign}\BibitemShut {NoStop}%
\bibitem [{\citenamefont {Perdew}\ \emph {et~al.}(2008)\citenamefont {Perdew},
  \citenamefont {Ruzsinszky}, \citenamefont {Csonka}, \citenamefont {Vydrov},
  \citenamefont {Scuseria}, \citenamefont {Constantin}, \citenamefont {Zhou},\
  and\ \citenamefont {Burke}}]{PhysRevLett.100.136406}%
  \BibitemOpen
  \bibfield  {author} {\bibinfo {author} {\bibfnamefont {J.~P.}\ \bibnamefont
  {Perdew}}, \bibinfo {author} {\bibfnamefont {A.}~\bibnamefont {Ruzsinszky}},
  \bibinfo {author} {\bibfnamefont {G.~I.}\ \bibnamefont {Csonka}}, \bibinfo
  {author} {\bibfnamefont {O.~A.}\ \bibnamefont {Vydrov}}, \bibinfo {author}
  {\bibfnamefont {G.~E.}\ \bibnamefont {Scuseria}}, \bibinfo {author}
  {\bibfnamefont {L.~A.}\ \bibnamefont {Constantin}}, \bibinfo {author}
  {\bibfnamefont {X.}~\bibnamefont {Zhou}}, \ and\ \bibinfo {author}
  {\bibfnamefont {K.}~\bibnamefont {Burke}},\ }\href {\doibase
  10.1103/PhysRevLett.100.136406} {\bibfield  {journal} {\bibinfo  {journal}
  {Phys. Rev. Lett.}\ }\textbf {\bibinfo {volume} {100}},\ \bibinfo {pages}
  {136406} (\bibinfo {year} {2008})}\BibitemShut {NoStop}%
\bibitem [{\citenamefont {Wannier}(1937)}]{PhysRev.52.191}%
  \BibitemOpen
  \bibfield  {author} {\bibinfo {author} {\bibfnamefont {G.~H.}\ \bibnamefont
  {Wannier}},\ }\href {\doibase 10.1103/PhysRev.52.191} {\bibfield  {journal}
  {\bibinfo  {journal} {Phys. Rev.}\ }\textbf {\bibinfo {volume} {52}},\
  \bibinfo {pages} {191} (\bibinfo {year} {1937})}\BibitemShut {NoStop}%
\bibitem [{\citenamefont {Marzari}\ \emph {et~al.}(2012)\citenamefont
  {Marzari}, \citenamefont {Mostofi}, \citenamefont {Yates}, \citenamefont
  {Souza},\ and\ \citenamefont {Vanderbilt}}]{RevModPhys.84.1419}%
  \BibitemOpen
  \bibfield  {author} {\bibinfo {author} {\bibfnamefont {N.}~\bibnamefont
  {Marzari}}, \bibinfo {author} {\bibfnamefont {A.~A.}\ \bibnamefont
  {Mostofi}}, \bibinfo {author} {\bibfnamefont {J.~R.}\ \bibnamefont {Yates}},
  \bibinfo {author} {\bibfnamefont {I.}~\bibnamefont {Souza}}, \ and\ \bibinfo
  {author} {\bibfnamefont {D.}~\bibnamefont {Vanderbilt}},\ }\href {\doibase
  10.1103/RevModPhys.84.1419} {\bibfield  {journal} {\bibinfo  {journal} {Rev.
  Mod. Phys.}\ }\textbf {\bibinfo {volume} {84}},\ \bibinfo {pages} {1419}
  (\bibinfo {year} {2012})}\BibitemShut {NoStop}%
\bibitem [{\citenamefont {Mostofi}\ \emph {et~al.}(2008)\citenamefont
  {Mostofi}, \citenamefont {Yates}, \citenamefont {Lee}, \citenamefont {Souza},
  \citenamefont {Vanderbilt},\ and\ \citenamefont
  {Marzari}}]{mostofi2008wannier90}%
  \BibitemOpen
  \bibfield  {author} {\bibinfo {author} {\bibfnamefont {A.~A.}\ \bibnamefont
  {Mostofi}}, \bibinfo {author} {\bibfnamefont {J.~R.}\ \bibnamefont {Yates}},
  \bibinfo {author} {\bibfnamefont {Y.-S.}\ \bibnamefont {Lee}}, \bibinfo
  {author} {\bibfnamefont {I.}~\bibnamefont {Souza}}, \bibinfo {author}
  {\bibfnamefont {D.}~\bibnamefont {Vanderbilt}}, \ and\ \bibinfo {author}
  {\bibfnamefont {N.}~\bibnamefont {Marzari}},\ }\href
  {https://doi.org/10.1016/j.cpc.2007.11.016} {\bibfield  {journal} {\bibinfo
  {journal} {Computer physics communications}\ }\textbf {\bibinfo {volume}
  {178}},\ \bibinfo {pages} {685} (\bibinfo {year} {2008})}\BibitemShut
  {NoStop}%
\bibitem [{\citenamefont {Kune{\v{s}}}\ \emph {et~al.}(2010)\citenamefont
  {Kune{\v{s}}}, \citenamefont {Arita}, \citenamefont {Wissgott}, \citenamefont
  {Toschi}, \citenamefont {Ikeda},\ and\ \citenamefont
  {Held}}]{kunevs2010wien2wannier}%
  \BibitemOpen
  \bibfield  {author} {\bibinfo {author} {\bibfnamefont {J.}~\bibnamefont
  {Kune{\v{s}}}}, \bibinfo {author} {\bibfnamefont {R.}~\bibnamefont {Arita}},
  \bibinfo {author} {\bibfnamefont {P.}~\bibnamefont {Wissgott}}, \bibinfo
  {author} {\bibfnamefont {A.}~\bibnamefont {Toschi}}, \bibinfo {author}
  {\bibfnamefont {H.}~\bibnamefont {Ikeda}}, \ and\ \bibinfo {author}
  {\bibfnamefont {K.}~\bibnamefont {Held}},\ }\href
  {https://doi.org/10.1016/j.cpc.2010.08.005} {\bibfield  {journal} {\bibinfo
  {journal} {Computer Physics Communications}\ }\textbf {\bibinfo {volume}
  {181}},\ \bibinfo {pages} {1888} (\bibinfo {year} {2010})}\BibitemShut
  {NoStop}%
\bibitem [{\citenamefont {Anisimov}\ \emph {et~al.}(1991)\citenamefont
  {Anisimov}, \citenamefont {Zaanen},\ and\ \citenamefont
  {Andersen}}]{Anisimov1991}%
  \BibitemOpen
  \bibfield  {author} {\bibinfo {author} {\bibfnamefont {V.~I.}\ \bibnamefont
  {Anisimov}}, \bibinfo {author} {\bibfnamefont {J.}~\bibnamefont {Zaanen}}, \
  and\ \bibinfo {author} {\bibfnamefont {O.~K.}\ \bibnamefont {Andersen}},\
  }\href {\doibase 10.1103/PhysRevB.44.943} {\bibfield  {journal} {\bibinfo
  {journal} {Phys. Rev. B}\ }\textbf {\bibinfo {volume} {44}},\ \bibinfo
  {pages} {943} (\bibinfo {year} {1991})}\BibitemShut {NoStop}%
\bibitem [{\citenamefont {Gull}\ \emph {et~al.}(2011)\citenamefont {Gull},
  \citenamefont {Millis}, \citenamefont {Lichtenstein}, \citenamefont
  {Rubtsov}, \citenamefont {Troyer},\ and\ \citenamefont
  {Werner}}]{RevModPhys.83.349}%
  \BibitemOpen
  \bibfield  {author} {\bibinfo {author} {\bibfnamefont {E.}~\bibnamefont
  {Gull}}, \bibinfo {author} {\bibfnamefont {A.~J.}\ \bibnamefont {Millis}},
  \bibinfo {author} {\bibfnamefont {A.~I.}\ \bibnamefont {Lichtenstein}},
  \bibinfo {author} {\bibfnamefont {A.~N.}\ \bibnamefont {Rubtsov}}, \bibinfo
  {author} {\bibfnamefont {M.}~\bibnamefont {Troyer}}, \ and\ \bibinfo {author}
  {\bibfnamefont {P.}~\bibnamefont {Werner}},\ }\href {\doibase
  10.1103/RevModPhys.83.349} {\bibfield  {journal} {\bibinfo  {journal} {Rev.
  Mod. Phys.}\ }\textbf {\bibinfo {volume} {83}},\ \bibinfo {pages} {349}
  (\bibinfo {year} {2011})}\BibitemShut {NoStop}%
\bibitem [{\citenamefont {Parragh}\ \emph {et~al.}(2012)\citenamefont
  {Parragh}, \citenamefont {Toschi}, \citenamefont {Held},\ and\ \citenamefont
  {Sangiovanni}}]{PhysRevB.86.155158}%
  \BibitemOpen
  \bibfield  {author} {\bibinfo {author} {\bibfnamefont {N.}~\bibnamefont
  {Parragh}}, \bibinfo {author} {\bibfnamefont {A.}~\bibnamefont {Toschi}},
  \bibinfo {author} {\bibfnamefont {K.}~\bibnamefont {Held}}, \ and\ \bibinfo
  {author} {\bibfnamefont {G.}~\bibnamefont {Sangiovanni}},\ }\href {\doibase
  10.1103/PhysRevB.86.155158} {\bibfield  {journal} {\bibinfo  {journal} {Phys.
  Rev. B}\ }\textbf {\bibinfo {volume} {86}},\ \bibinfo {pages} {155158}
  (\bibinfo {year} {2012})}\BibitemShut {NoStop}%
\bibitem [{\citenamefont {Wallerberger}\ \emph {et~al.}(2019)\citenamefont
  {Wallerberger}, \citenamefont {Hausoel}, \citenamefont {Gunacker},
  \citenamefont {Kowalski}, \citenamefont {Parragh}, \citenamefont {Goth},
  \citenamefont {Held},\ and\ \citenamefont
  {Sangiovanni}}]{wallerberger2019w2dynamics}%
  \BibitemOpen
  \bibfield  {author} {\bibinfo {author} {\bibfnamefont {M.}~\bibnamefont
  {Wallerberger}}, \bibinfo {author} {\bibfnamefont {A.}~\bibnamefont
  {Hausoel}}, \bibinfo {author} {\bibfnamefont {P.}~\bibnamefont {Gunacker}},
  \bibinfo {author} {\bibfnamefont {A.}~\bibnamefont {Kowalski}}, \bibinfo
  {author} {\bibfnamefont {N.}~\bibnamefont {Parragh}}, \bibinfo {author}
  {\bibfnamefont {F.}~\bibnamefont {Goth}}, \bibinfo {author} {\bibfnamefont
  {K.}~\bibnamefont {Held}}, \ and\ \bibinfo {author} {\bibfnamefont
  {G.}~\bibnamefont {Sangiovanni}},\ }\href
  {https://doi.org/10.1016/j.cpc.2018.09.007} {\bibfield  {journal} {\bibinfo
  {journal} {Computer Physics Communications}\ }\textbf {\bibinfo {volume}
  {235}},\ \bibinfo {pages} {388} (\bibinfo {year} {2019})}\BibitemShut
  {NoStop}%
\bibitem [{\citenamefont {{Kaufmann}}\ and\ \citenamefont
  {{Held}}(2021)}]{Kaufmann2021}%
  \BibitemOpen
  \bibfield  {author} {\bibinfo {author} {\bibfnamefont {J.}~\bibnamefont
  {{Kaufmann}}}\ and\ \bibinfo {author} {\bibfnamefont {K.}~\bibnamefont
  {{Held}}},\ }\href {https://doi.org/10.48550/arXiv.2105.11211} {\bibfield
  {journal} {\bibinfo  {journal} {arXiv:2105.11211}\ } (\bibinfo {year}
  {2021})}\BibitemShut {NoStop}%
\bibitem [{\citenamefont {Gubernatis}\ \emph {et~al.}(1991)\citenamefont
  {Gubernatis}, \citenamefont {Jarrell}, \citenamefont {Silver},\ and\
  \citenamefont {Sivia}}]{PhysRevB.44.6011}%
  \BibitemOpen
  \bibfield  {author} {\bibinfo {author} {\bibfnamefont {J.~E.}\ \bibnamefont
  {Gubernatis}}, \bibinfo {author} {\bibfnamefont {M.}~\bibnamefont {Jarrell}},
  \bibinfo {author} {\bibfnamefont {R.~N.}\ \bibnamefont {Silver}}, \ and\
  \bibinfo {author} {\bibfnamefont {D.~S.}\ \bibnamefont {Sivia}},\ }\href
  {\doibase 10.1103/PhysRevB.44.6011} {\bibfield  {journal} {\bibinfo
  {journal} {Phys. Rev. B}\ }\textbf {\bibinfo {volume} {44}},\ \bibinfo
  {pages} {6011} (\bibinfo {year} {1991})}\BibitemShut {NoStop}%
\bibitem [{\citenamefont {Sandvik}(1998)}]{PhysRevB.57.10287}%
  \BibitemOpen
  \bibfield  {author} {\bibinfo {author} {\bibfnamefont {A.~W.}\ \bibnamefont
  {Sandvik}},\ }\href {\doibase 10.1103/PhysRevB.57.10287} {\bibfield
  {journal} {\bibinfo  {journal} {Phys. Rev. B}\ }\textbf {\bibinfo {volume}
  {57}},\ \bibinfo {pages} {10287} (\bibinfo {year} {1998})}\BibitemShut
  {NoStop}%
\bibitem [{\citenamefont {Liechtenstein}\ \emph {et~al.}(1995)\citenamefont
  {Liechtenstein}, \citenamefont {Anisimov},\ and\ \citenamefont
  {Zaanen}}]{Liechtenstein_1995}%
  \BibitemOpen
  \bibfield  {author} {\bibinfo {author} {\bibfnamefont {A.~I.}\ \bibnamefont
  {Liechtenstein}}, \bibinfo {author} {\bibfnamefont {V.~I.}\ \bibnamefont
  {Anisimov}}, \ and\ \bibinfo {author} {\bibfnamefont {J.}~\bibnamefont
  {Zaanen}},\ }\href {\doibase 10.1103/PhysRevB.52.R5467} {\bibfield  {journal}
  {\bibinfo  {journal} {Phys. Rev. B}\ }\textbf {\bibinfo {volume} {52}},\
  \bibinfo {pages} {R5467} (\bibinfo {year} {1995})}\BibitemShut {NoStop}%
\bibitem [{\citenamefont {Perdew}\ \emph {et~al.}(1996)\citenamefont {Perdew},
  \citenamefont {Burke},\ and\ \citenamefont
  {Ernzerhof}}]{perdew_generalized_1996}%
  \BibitemOpen
  \bibfield  {author} {\bibinfo {author} {\bibfnamefont {J.~P.}\ \bibnamefont
  {Perdew}}, \bibinfo {author} {\bibfnamefont {K.}~\bibnamefont {Burke}}, \
  and\ \bibinfo {author} {\bibfnamefont {M.}~\bibnamefont {Ernzerhof}},\ }\href
  {\doibase 10.1103/PhysRevLett.77.3865} {\bibfield  {journal} {\bibinfo
  {journal} {Physical Review Letters}\ }\textbf {\bibinfo {volume} {77}},\
  \bibinfo {pages} {3865} (\bibinfo {year} {1996})},\ \bibinfo {note}
  {publisher: American Physical Society}\BibitemShut {NoStop}%
\bibitem [{\citenamefont {Perdew}\ \emph {et~al.}(1997)\citenamefont {Perdew},
  \citenamefont {Burke},\ and\ \citenamefont
  {Ernzerhof}}]{perdew_generalized_1997}%
  \BibitemOpen
  \bibfield  {author} {\bibinfo {author} {\bibfnamefont {J.~P.}\ \bibnamefont
  {Perdew}}, \bibinfo {author} {\bibfnamefont {K.}~\bibnamefont {Burke}}, \
  and\ \bibinfo {author} {\bibfnamefont {M.}~\bibnamefont {Ernzerhof}},\ }\href
  {\doibase 10.1103/PhysRevLett.78.1396} {\bibfield  {journal} {\bibinfo
  {journal} {Physical Review Letters}\ }\textbf {\bibinfo {volume} {78}},\
  \bibinfo {pages} {1396} (\bibinfo {year} {1997})},\ \bibinfo {note}
  {publisher: American Physical Society}\BibitemShut {NoStop}%
\bibitem [{\citenamefont {Hicks}\ \emph {et~al.}(2018)\citenamefont {Hicks},
  \citenamefont {Oses}, \citenamefont {Gossett}, \citenamefont {Gomez},
  \citenamefont {Taylor}, \citenamefont {Toher}, \citenamefont {Mehl},
  \citenamefont {Levy},\ and\ \citenamefont {Curtarolo}}]{Hicks_2018}%
  \BibitemOpen
  \bibfield  {author} {\bibinfo {author} {\bibfnamefont {D.}~\bibnamefont
  {Hicks}}, \bibinfo {author} {\bibfnamefont {C.}~\bibnamefont {Oses}},
  \bibinfo {author} {\bibfnamefont {E.}~\bibnamefont {Gossett}}, \bibinfo
  {author} {\bibfnamefont {G.}~\bibnamefont {Gomez}}, \bibinfo {author}
  {\bibfnamefont {R.~H.}\ \bibnamefont {Taylor}}, \bibinfo {author}
  {\bibfnamefont {C.}~\bibnamefont {Toher}}, \bibinfo {author} {\bibfnamefont
  {M.~J.}\ \bibnamefont {Mehl}}, \bibinfo {author} {\bibfnamefont
  {O.}~\bibnamefont {Levy}}, \ and\ \bibinfo {author} {\bibfnamefont
  {S.}~\bibnamefont {Curtarolo}},\ }\href
  {https://doi.org/10.1107/S2053273318003066} {\bibfield  {journal} {\bibinfo
  {journal} {Acta Crystallographica Section A}\ }\textbf {\bibinfo {volume}
  {74}},\ \bibinfo {pages} {184} (\bibinfo {year} {2018})}\BibitemShut
  {NoStop}%
\bibitem [{\citenamefont {{Kitatani}}\ \emph {et~al.}(2020)\citenamefont
  {{Kitatani}}, \citenamefont {{Si}}, \citenamefont {{Janson}}, \citenamefont
  {{Arita}}, \citenamefont {{Zhong}},\ and\ \citenamefont
  {{Held}}}]{Kitatani2020}%
  \BibitemOpen
  \bibfield  {author} {\bibinfo {author} {\bibfnamefont {M.}~\bibnamefont
  {{Kitatani}}}, \bibinfo {author} {\bibfnamefont {L.}~\bibnamefont {{Si}}},
  \bibinfo {author} {\bibfnamefont {O.}~\bibnamefont {{Janson}}}, \bibinfo
  {author} {\bibfnamefont {R.}~\bibnamefont {{Arita}}}, \bibinfo {author}
  {\bibfnamefont {Z.}~\bibnamefont {{Zhong}}}, \ and\ \bibinfo {author}
  {\bibfnamefont {K.}~\bibnamefont {{Held}}},\ }\href {\doibase
  10.1038/s41535-020-00260-y} {\bibfield  {journal} {\bibinfo  {journal} {npj
  Quantum Materials}\ }\textbf {\bibinfo {volume} {5}},\ \bibinfo {pages} {59}
  (\bibinfo {year} {2020})}\BibitemShut {NoStop}%
\bibitem [{\citenamefont {De~Visser}\ \emph {et~al.}(1987)\citenamefont
  {De~Visser}, \citenamefont {Menovsky},\ and\ \citenamefont
  {Franse}}]{de1987upt3}%
  \BibitemOpen
  \bibfield  {author} {\bibinfo {author} {\bibfnamefont {A.}~\bibnamefont
  {De~Visser}}, \bibinfo {author} {\bibfnamefont {A.}~\bibnamefont {Menovsky}},
  \ and\ \bibinfo {author} {\bibfnamefont {J.}~\bibnamefont {Franse}},\ }\href
  {\doibase 10.1016/0378-4363(87)90008-8} {\bibfield  {journal} {\bibinfo
  {journal} {Physica B+C}\ }\textbf {\bibinfo {volume} {147}},\ \bibinfo
  {pages} {81} (\bibinfo {year} {1987})}\BibitemShut {NoStop}%
\bibitem [{\citenamefont {Sauls}(1994)}]{sauls1994order}%
  \BibitemOpen
  \bibfield  {author} {\bibinfo {author} {\bibfnamefont {J.}~\bibnamefont
  {Sauls}},\ }\href {\doibase 10.1080/00018739400101475} {\bibfield  {journal}
  {\bibinfo  {journal} {Advances in Physics}\ }\textbf {\bibinfo {volume}
  {43}},\ \bibinfo {pages} {113} (\bibinfo {year} {1994})}\BibitemShut
  {NoStop}%
\bibitem [{\citenamefont {Li}\ \emph {et~al.}(2022)\citenamefont {Li},
  \citenamefont {Tong}, \citenamefont {Shao}, \citenamefont {Bao},
  \citenamefont {Frauenheim},\ and\ \citenamefont {Liu}}]{li2022anomalously}%
  \BibitemOpen
  \bibfield  {author} {\bibinfo {author} {\bibfnamefont {S.}~\bibnamefont
  {Li}}, \bibinfo {author} {\bibfnamefont {Z.}~\bibnamefont {Tong}}, \bibinfo
  {author} {\bibfnamefont {C.}~\bibnamefont {Shao}}, \bibinfo {author}
  {\bibfnamefont {H.}~\bibnamefont {Bao}}, \bibinfo {author} {\bibfnamefont
  {T.}~\bibnamefont {Frauenheim}}, \ and\ \bibinfo {author} {\bibfnamefont
  {X.}~\bibnamefont {Liu}},\ }\href {\doibase 10.1021/acs.jpclett.2c00425}
  {\bibfield  {journal} {\bibinfo  {journal} {The Journal of Physical Chemistry
  Letters}\ }\textbf {\bibinfo {volume} {13}},\ \bibinfo {pages} {4289}
  (\bibinfo {year} {2022})}\BibitemShut {NoStop}%
\bibitem [{\citenamefont {Miyake}\ and\ \citenamefont
  {Aryasetiawan}(2008)}]{PhysRevB.77.085122}%
  \BibitemOpen
  \bibfield  {author} {\bibinfo {author} {\bibfnamefont {T.}~\bibnamefont
  {Miyake}}\ and\ \bibinfo {author} {\bibfnamefont {F.}~\bibnamefont
  {Aryasetiawan}},\ }\href {\doibase 10.1103/PhysRevB.77.085122} {\bibfield
  {journal} {\bibinfo  {journal} {Phys. Rev. B}\ }\textbf {\bibinfo {volume}
  {77}},\ \bibinfo {pages} {085122} (\bibinfo {year} {2008})}\BibitemShut
  {NoStop}%
\bibitem [{\citenamefont {Jang}\ \emph {et~al.}(2016)\citenamefont {Jang},
  \citenamefont {Sakakibara}, \citenamefont {Kino}, \citenamefont {Kotani},
  \citenamefont {Kuroki},\ and\ \citenamefont {Han}}]{Jang2016}%
  \BibitemOpen
  \bibfield  {author} {\bibinfo {author} {\bibfnamefont {S.~W.}\ \bibnamefont
  {Jang}}, \bibinfo {author} {\bibfnamefont {H.}~\bibnamefont {Sakakibara}},
  \bibinfo {author} {\bibfnamefont {H.}~\bibnamefont {Kino}}, \bibinfo {author}
  {\bibfnamefont {T.}~\bibnamefont {Kotani}}, \bibinfo {author} {\bibfnamefont
  {K.}~\bibnamefont {Kuroki}}, \ and\ \bibinfo {author} {\bibfnamefont {M.~J.}\
  \bibnamefont {Han}},\ }\href {\doibase 10.1038/srep33397} {\bibfield
  {journal} {\bibinfo  {journal} {Scientific Reports}\ }\textbf {\bibinfo
  {volume} {6}},\ \bibinfo {pages} {33397} (\bibinfo {year}
  {2016})}\BibitemShut {NoStop}%
\bibitem [{\citenamefont {Di~Sante}\ \emph {et~al.}(2017)\citenamefont
  {Di~Sante}, \citenamefont {Hausoel}, \citenamefont {Barone}, \citenamefont
  {Tomczak}, \citenamefont {Sangiovanni},\ and\ \citenamefont
  {Thomale}}]{DiSante_cubio}%
  \BibitemOpen
  \bibfield  {author} {\bibinfo {author} {\bibfnamefont {D.}~\bibnamefont
  {Di~Sante}}, \bibinfo {author} {\bibfnamefont {A.}~\bibnamefont {Hausoel}},
  \bibinfo {author} {\bibfnamefont {P.}~\bibnamefont {Barone}}, \bibinfo
  {author} {\bibfnamefont {J.~M.}\ \bibnamefont {Tomczak}}, \bibinfo {author}
  {\bibfnamefont {G.}~\bibnamefont {Sangiovanni}}, \ and\ \bibinfo {author}
  {\bibfnamefont {R.}~\bibnamefont {Thomale}},\ }\href {\doibase
  10.1103/PhysRevB.96.121106} {\bibfield  {journal} {\bibinfo  {journal} {Phys.
  Rev. B}\ }\textbf {\bibinfo {volume} {96}},\ \bibinfo {pages} {121106}
  (\bibinfo {year} {2017})}\BibitemShut {NoStop}%
\bibitem [{\citenamefont {Ivashko}\ \emph {et~al.}(2019)\citenamefont
  {Ivashko}, \citenamefont {Horio}, \citenamefont {Wan}, \citenamefont
  {Christensen}, \citenamefont {McNally}, \citenamefont {Paris}, \citenamefont
  {Tseng}, \citenamefont {Shaik}, \citenamefont {R{\o}nnow}, \citenamefont
  {Wei}, \citenamefont {Adamo}, \citenamefont {Lichtensteiger}, \citenamefont
  {Gibert}, \citenamefont {Beasley}, \citenamefont {Shen}, \citenamefont
  {Tomczak}, \citenamefont {Schmitt},\ and\ \citenamefont
  {Chang}}]{Chang_LSCO}%
  \BibitemOpen
  \bibfield  {author} {\bibinfo {author} {\bibfnamefont {O.}~\bibnamefont
  {Ivashko}}, \bibinfo {author} {\bibfnamefont {M.}~\bibnamefont {Horio}},
  \bibinfo {author} {\bibfnamefont {W.}~\bibnamefont {Wan}}, \bibinfo {author}
  {\bibfnamefont {N.~B.}\ \bibnamefont {Christensen}}, \bibinfo {author}
  {\bibfnamefont {D.~E.}\ \bibnamefont {McNally}}, \bibinfo {author}
  {\bibfnamefont {E.}~\bibnamefont {Paris}}, \bibinfo {author} {\bibfnamefont
  {Y.}~\bibnamefont {Tseng}}, \bibinfo {author} {\bibfnamefont {N.~E.}\
  \bibnamefont {Shaik}}, \bibinfo {author} {\bibfnamefont {H.~M.}\ \bibnamefont
  {R{\o}nnow}}, \bibinfo {author} {\bibfnamefont {H.~I.}\ \bibnamefont {Wei}},
  \bibinfo {author} {\bibfnamefont {C.}~\bibnamefont {Adamo}}, \bibinfo
  {author} {\bibfnamefont {C.}~\bibnamefont {Lichtensteiger}}, \bibinfo
  {author} {\bibfnamefont {M.}~\bibnamefont {Gibert}}, \bibinfo {author}
  {\bibfnamefont {M.~R.}\ \bibnamefont {Beasley}}, \bibinfo {author}
  {\bibfnamefont {K.~M.}\ \bibnamefont {Shen}}, \bibinfo {author}
  {\bibfnamefont {J.~M.}\ \bibnamefont {Tomczak}}, \bibinfo {author}
  {\bibfnamefont {T.}~\bibnamefont {Schmitt}}, \ and\ \bibinfo {author}
  {\bibfnamefont {J.}~\bibnamefont {Chang}},\ }\href {\doibase
  10.1038/s41467-019-08664-6} {\bibfield  {journal} {\bibinfo  {journal}
  {Nature Communications}\ }\textbf {\bibinfo {volume} {10}},\ \bibinfo {pages}
  {786} (\bibinfo {year} {2019})}\BibitemShut {NoStop}%
\bibitem [{\citenamefont {Nilsson}\ \emph {et~al.}(2019)\citenamefont
  {Nilsson}, \citenamefont {Karlsson},\ and\ \citenamefont
  {Aryasetiawan}}]{PhysRevB.99.075135}%
  \BibitemOpen
  \bibfield  {author} {\bibinfo {author} {\bibfnamefont {F.}~\bibnamefont
  {Nilsson}}, \bibinfo {author} {\bibfnamefont {K.}~\bibnamefont {Karlsson}}, \
  and\ \bibinfo {author} {\bibfnamefont {F.}~\bibnamefont {Aryasetiawan}},\
  }\href {\doibase 10.1103/PhysRevB.99.075135} {\bibfield  {journal} {\bibinfo
  {journal} {Phys. Rev. B}\ }\textbf {\bibinfo {volume} {99}},\ \bibinfo
  {pages} {075135} (\bibinfo {year} {2019})}\BibitemShut {NoStop}%
\bibitem [{\citenamefont {Hansmann}\ \emph {et~al.}(2014)\citenamefont
  {Hansmann}, \citenamefont {Parragh}, \citenamefont {Toschi}, \citenamefont
  {Sangiovanni},\ and\ \citenamefont {Held}}]{Hansmann2014}%
  \BibitemOpen
  \bibfield  {author} {\bibinfo {author} {\bibfnamefont {P.}~\bibnamefont
  {Hansmann}}, \bibinfo {author} {\bibfnamefont {N.}~\bibnamefont {Parragh}},
  \bibinfo {author} {\bibfnamefont {A.}~\bibnamefont {Toschi}}, \bibinfo
  {author} {\bibfnamefont {G.}~\bibnamefont {Sangiovanni}}, \ and\ \bibinfo
  {author} {\bibfnamefont {K.}~\bibnamefont {Held}},\ }\href
  {http://stacks.iop.org/1367-2630/16/i=3/a=033009} {\bibfield  {journal}
  {\bibinfo  {journal} {New J. Phys.}\ }\textbf {\bibinfo {volume} {16}},\
  \bibinfo {pages} {033009} (\bibinfo {year} {2014})}\BibitemShut {NoStop}%
\bibitem [{\citenamefont {Weber}\ \emph {et~al.}(2012)\citenamefont {Weber},
  \citenamefont {Yee}, \citenamefont {Haule},\ and\ \citenamefont
  {Kotliar}}]{Weber_2012}%
  \BibitemOpen
  \bibfield  {author} {\bibinfo {author} {\bibfnamefont {C.}~\bibnamefont
  {Weber}}, \bibinfo {author} {\bibfnamefont {C.}~\bibnamefont {Yee}}, \bibinfo
  {author} {\bibfnamefont {K.}~\bibnamefont {Haule}}, \ and\ \bibinfo {author}
  {\bibfnamefont {G.}~\bibnamefont {Kotliar}},\ }\href {\doibase
  10.1209/0295-5075/100/37001} {\bibfield  {journal} {\bibinfo  {journal}
  {{EPL} (Europhysics Letters)}\ }\textbf {\bibinfo {volume} {100}},\ \bibinfo
  {pages} {37001} (\bibinfo {year} {2012})}\BibitemShut {NoStop}%
\bibitem [{\citenamefont {Yang}\ \emph {et~al.}(2017)\citenamefont {Yang},
  \citenamefont {Sobota}, \citenamefont {He}, \citenamefont {Wang},
  \citenamefont {Leuenberger}, \citenamefont {Soifer}, \citenamefont
  {Hashimoto}, \citenamefont {Lu}, \citenamefont {Eisaki}, \citenamefont
  {Moritz}, \citenamefont {Devereaux}, \citenamefont {Kirchmann},\ and\
  \citenamefont {Shen}}]{Yang2017}%
  \BibitemOpen
  \bibfield  {author} {\bibinfo {author} {\bibfnamefont {S.-L.}\ \bibnamefont
  {Yang}}, \bibinfo {author} {\bibfnamefont {J.~A.}\ \bibnamefont {Sobota}},
  \bibinfo {author} {\bibfnamefont {Y.}~\bibnamefont {He}}, \bibinfo {author}
  {\bibfnamefont {Y.}~\bibnamefont {Wang}}, \bibinfo {author} {\bibfnamefont
  {D.}~\bibnamefont {Leuenberger}}, \bibinfo {author} {\bibfnamefont
  {H.}~\bibnamefont {Soifer}}, \bibinfo {author} {\bibfnamefont
  {M.}~\bibnamefont {Hashimoto}}, \bibinfo {author} {\bibfnamefont {D.~H.}\
  \bibnamefont {Lu}}, \bibinfo {author} {\bibfnamefont {H.}~\bibnamefont
  {Eisaki}}, \bibinfo {author} {\bibfnamefont {B.}~\bibnamefont {Moritz}},
  \bibinfo {author} {\bibfnamefont {T.~P.}\ \bibnamefont {Devereaux}}, \bibinfo
  {author} {\bibfnamefont {P.~S.}\ \bibnamefont {Kirchmann}}, \ and\ \bibinfo
  {author} {\bibfnamefont {Z.-X.}\ \bibnamefont {Shen}},\ }\href {\doibase
  10.1103/PhysRevB.96.245112} {\bibfield  {journal} {\bibinfo  {journal} {Phys.
  Rev. B}\ }\textbf {\bibinfo {volume} {96}},\ \bibinfo {pages} {245112}
  (\bibinfo {year} {2017})}\BibitemShut {NoStop}%
\end{thebibliography}
\end{document}